\documentclass[12pt,preprint]{aastex}

\newcommand\vi{\mbox{$V\!-\!I$}}
\newcommand{\Sec}{${}^{\prime\prime}$\llap{.}}
\newcommand{\Min}{${}^{\prime}$\llap{.}}

\title{The Luminosity Function and Color-Magnitude Diagram of the Globular
Cluster M12} 
\author{Jonathan R. Hargis and Eric L. Sandquist}
\affil{Department of Astronomy, San Diego State University,
5500 Campanile Drive, San Diego, CA 92182}
\email{jhargis@sciences.sdsu.edu, erics@mintaka.sdsu.edu}

\and

\author{Michael Bolte}
\affil{University of California Observatories/Lick Observatory, University
of California, Santa Cruz, CA 95064}
\email{bolte@ucolick.org}

\shorttitle{Luminosity Function of M12}
\shortauthors{Hargis, Sandquist, & Bolte}

\begin{document}

\begin{abstract}
In this paper we present the $V$ and $I$ luminosity functions
and color-magnitude diagrams derived from wide-field
($23\arcmin\times23\arcmin$) $BVI$ photometry of the intermediate
metallicity ([Fe/H]$\sim-1.3$) Galactic globular cluster M12. Using
observed values (and ranges of values) for the cluster metallicity,
reddening, distance modulus, and age we compare these data to recent
$\alpha$-enhanced stellar evolution models for low mass metal-poor
stars. We describe several methods of making comparisons between
theoretical and observed luminosity functions in order to isolate
the evolutionary
timescale information the luminosity functions contain. We find no
significant evidence
of excesses of stars on the red giant branch, although the morphology
of the subgiant branch in the observed luminosity function does
not match theoretical
predictions in a satisfactory way. Current uncertainties in
$T_{\mbox{eff}}$-color transformations (and possibly also in other
physics inputs to the models) make more detailed conclusions about the
subgiant branch morphology impossible. Given the recent constraints on
cluster ages from the WMAP experiment (Spergel et al. 2003), we find
that good fitting models that do not include He diffusion (both
color-magnitude diagrams and luminosity functions) are too old (by
$\sim1-2$ Gyr) to adequately represent the cluster luminosity function.
The inclusion of helium diffusion in the models provides an age reduction
(compared to non-diffusive models) that is consistent with the age of the
universe being $13.7\pm0.2$ Gyr (Bennett et al. 2003).
\end{abstract}


\section{Introduction}
One of the observational tools available for the study of low-mass
($\sim0.5-1~M_{\odot}$), metal-poor stars is the luminosity function
(LF) of Galactic globular clusters (GGCs).  A GGC usually presents a
large, chemically homogeneous, and coeval stellar population ---
samples of a kind that cannot easily be extracted from halo field
stars. In particular, the LF counts of evolved stars (from the
main-sequence turnoff to the tip of the red giant branch)
are directly related to the rate of nuclear fuel consumption
\citep{re88}.  Thus, a high-precision LF indirectly reflects interior
stellar physics, and thus can complement observations of
stellar surface conditions.

Previous LF studies of GGCs have 
given hints that non-standard physics might
be required in the models.  First, in the case of several metal-poor clusters an
excess of stars on the sub-giant branch (SGB) has been noted.  In a
study that combined the LFs of M68, NGC 6397, and M92, \citet{st91}
found an excess located just brighter than the main-sequence turnoff
(MSTO).  A similar excess was also observed in M30 by \citet{bo94} and
confirmed by \citet{be96}, \citet{gu98}, and \citet{sa99}.  Such an
excess could be the result of enhanced energy transport in the cores
of main-sequence (MS) stars nearing hydrogen exhaustion (e.g.
Faulkner \& Swenson 1993).  To date excesses have only been seen in
extremely metal-poor clusters and at low statistical significance; the LFs
of M5 \citep{sand96} and M3 \citep{rood99} did not reveal such excesses.
Second, studies by \citet{st91}, \citet{bv92}, \citet{bo94}, \citet{be96},
and \citet{sa99} have shown that there may exist an over-abundance of red
giant branch (RGB) stars relative to the number of MS stars in some
clusters. This may be indicative of physical processes such as core rotation 
\citep{va98} and/or deep mixing \citep{la00}.  The observed LFs of other
GGCs, however, have been argued to agree with theory \citep{de97,zo00,rood99}.
While the LF of M3 derived by \citet{rood99} is one of the largest to date,
the study by \citet{zo00} has a smaller overall statistical significance due to 
the smaller overall star samples.  In general, only thorough studies of
large star populations in GGCs will help confirm or deny the reality of
these kinds of excesses.

A physical effect that is becoming part of standard stellar
models is the gravitational settling of heavy elements \citep{ri02}.
Solar models show \citep{pr94,ri96,ba97,gu97} that helioseismic
data and the inferred radius of the convection zone \citep{cgt91} can
only be matched by theory if the surface abundance of helium has
decreased with time.  Stellar evolution models for low mass,
metal-poor stars that include the effects of He diffusion have been
previously constructed \citep{pr91,ch92,st97,ch01} to investigate the
observational effects on GGC LFs and color-magnitude diagrams (CMDs).
One important result of these studies was that by adding He to the
core of the stars, and consequently displacing H, the duration of the
MS lifetime is shortened.  This results in a lower MSTO luminosity for a 
given age and has strong implications for the
derivation of GGC ages \citep{ch92,va02}.  Using
these diffusive models GGC ages may be reduced as much as $10-15$\%
($\sim1-2$ Gyr) compared to models that do not incorporate diffusion
\citep{st97,va02}, although the reduction could be as little as $4-7$\%
($\sim0.5-1$ Gyr) depending on the presence of complete or partial
ionization in the models \citep{ri02,gr03}. Given the recent analysis of
the WMAP data from the cosmic background  radiation, we now have a tight
cosmological upper-limit on the age of the universe of $13.7\pm0.2$ Gyr
\citep{be03,sp03}.  As a consequence, theoretical stellar evolution
models that do not include diffusion (at least for He) may produce
globular cluster isochrones that are too old.

In this paper we present $BVI$ photometric data derived from
wide-field CCD photometry of the GGC M12 (NGC 6218, C 1644-018).  This
bright, large, intermediate metallicity ([Fe/H]$\sim1.3$) cluster should
provide an interesting comparison to other well-studied intermediate 
clusters such as M3 and M5.  M12 is also an extreme ``second parameter'' 
cluster [$(B-R)/(B+V+R)=0.92$; \citet{le94}]. Ground-based photometric data
on M12 have most recently been presented by von Braun et al. (2002;
hereafter VB02), Rosenberg et al. (2000; hereafter R00), and Brocato et al.
(1996; hereafter B96). Space-based data on M12 was published by
\citet{pi02} as part of a \textit{HST} GGC snapshot survey. Sato, Richer, \& 
Fahlman (1989; hereafter S89) presented a deep CMD and LF for the inner 
regions of M12, but to date no LF of the evolved stellar populations of M12 
has been determined.

In this study we focus on comparing our data (in the form of LFs and
CMDs) to three sets of stellar evolution models.  In $\S2$ we discuss
the observations, data reduction and photometry, photometric
calibration, and comparisons to existing photometry. In $\S3$, we
present the photometry in the form of CMDs and derived fiducial lines.
$\S4$ discusses the cluster reddening, distance modulus, age, and
metallicity which are the necessary input parameters to the comparison
of the theoretical LF with the observed.  The computation of the
observed LF and incompleteness corrections are discussed in $\S5$.  In
$\S6$ we compare the data to theoretical CMDs and LFs.  Our conclusions 
are presented in $\S7$.

\section{Observations}

Observations for this study were done on the nights of UT date 6 May
1995 and 9 May 1995 using the Kitt Peak National Observatory (KPNO)
0.9 m telescope.  In total, 12 images were obtained in $BVI$
filters (four images per filter).  Three images in each band were
taken on night 3 (6 May 1995) of the run, with exposure times of 10,
60 and 200 s. One additional 60 s image in each filter was obtained
on night 6 (9 May 1995) of the run.  Seeing
conditions were approximately 1\Sec5 on night 3 of the run and
$2\arcsec$ on night 6 of the run.  All data were taken using a
$2048\times2048$ pixel CCD chip with a plate scale of 0\Sec68
pixel$^{-1}$, so that the total sky coverage was
23\Min2$\times$23\Min2 around the cluster center.

\subsection{Data Reduction}

The frames were reduced in the standard fashion using
IRAF\footnote{IRAF (Image Reduction and Analysis Facility) is
distributed by the National Optical Astronomy Observatories, which are
operated by the Association of Universities for Research in Astronomy,
Inc., under contract with the National Science Foundation.} tasks and
packages. The bias level was removed by subtracting fits to the overscan
region and a master `zero' frame. Both twilight and dome flat fields
were used in constructing a master flat field
frame from the high spatial frequency component of the
dome flats and the low-frequency (smoothed) component of the twilight
flats.

\subsection{Object Frames}

The M12 profile-fitting photometry was performed using the DAOPHOT
II/ALLSTAR package of programs \citep{pbs87}.  In general, about 120
stars were used to determine the point-spread function (PSF) in each
frame. Stars were rejected as candidates for the PSF determination if
the FWHM of their profile varied by more than 3$\sigma$ from the mean.
The radial profiles of the remaining candidate stars were then examined
individually to reject any stars that had nearby, faint companions.

In order to obtain a master list of stars for each frame, an iterative
procedure using DAOPHOT's FIND routine and ALLSTAR was implemented.
The final list of 17,303 stars in this study was determined from the
master star lists of the three filters. This master list was used as
an input to a final run of ALLSTAR to determine photometry from a
consistent list of stars.  The use of the ALLFRAME package \citep{pbs94}
did not provide a noticeable improvement of the photometry.

\subsection{Calibration Against Primary Standards}

Observations of Landolt standard fields and a number of cluster fields
were made on night 6 of the run under photometric conditions. The
standard star fields were observed at a range of airmasses in order to
determine atmospheric extinction coefficients. We have chosen to use
standard values from the extensive tabulation of Stetson (2000) for
the calibration because those standard stars have been shown to be
accurately on the same photometric scale as the earlier Landolt (1992)
tabulation and because there is a large number of standard
stars covering a larger range of colors.

\begin{figure}
\plotone{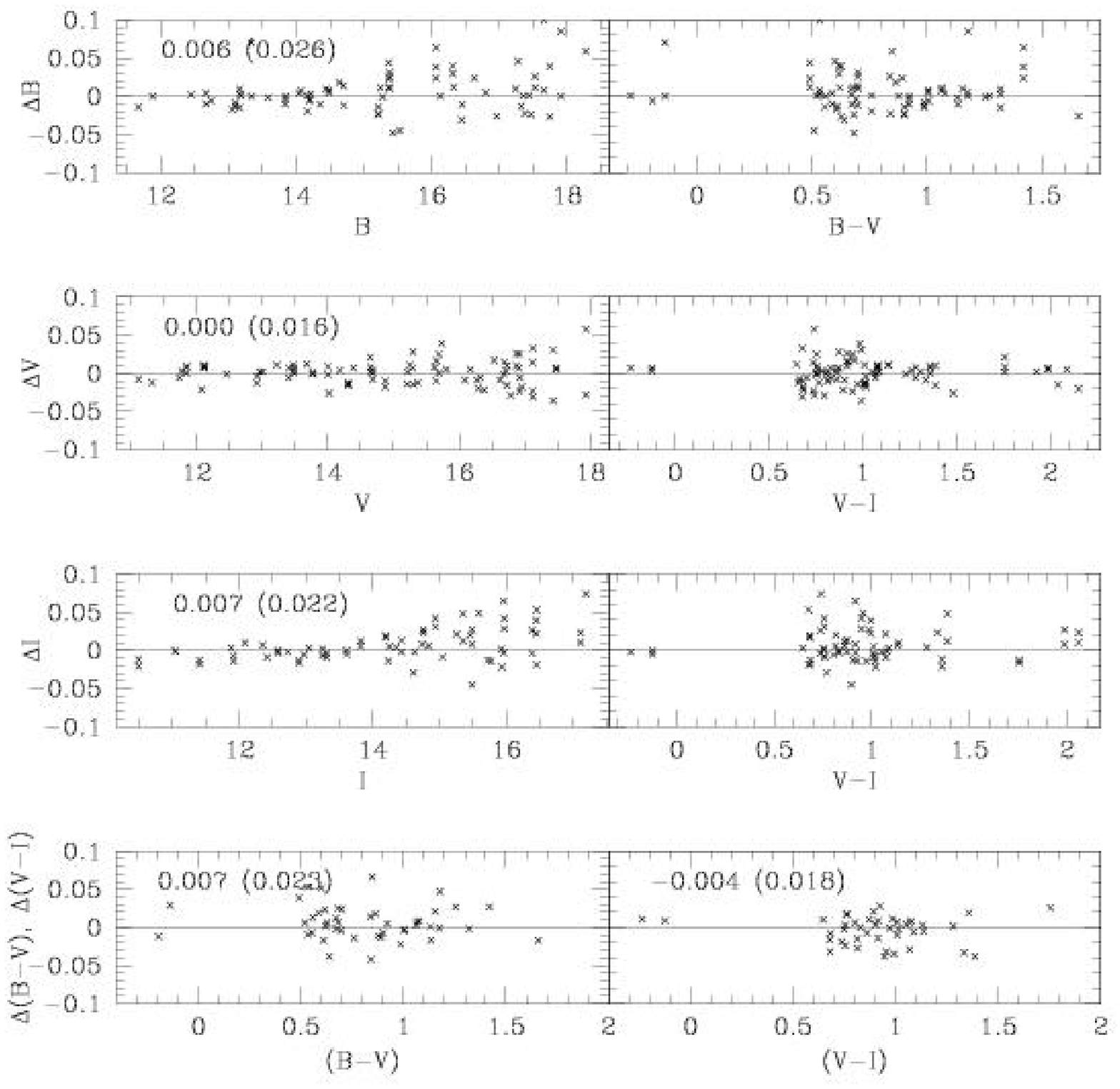}
\caption{Photometric residuals [in the sense of our values minus those
of \citet{land92} and \citet{pbs00}] from the calibration of primary
standard stars.  The RMS residuals are listed in the panels
(with the standard deviations given in parentheses).  The number of stars
in the respective plots are 75 in $\Delta B$, 89 in $\Delta V$, and 68 in
$\Delta I$.
\label{primary}}
\end{figure}

Aperture photometry was performed on both standard and cluster frames
using DAOPHOT II using multiple synthetic
apertures. Growth curves were used to extrapolate measurements to a
(large) common aperture size using the program DAOGROW (Stetson 1990).
The photometric transformation equations used in the calibration were
\[ b = B + a_0 + (-0.0686\pm0.0052)(\bv) + (0.2537\pm0.0139)(X - 1.25)\]
\[ v = V + b_0 + (0.0225\pm0.0033)(\vi) +  (0.1810\pm0.0090)(X - 1.25)\]
\[ i = I + c_0 + (-0.0023\pm0.0047)(\vi) + (0.1285\pm0.0132)(X - 1.25)\]
where $b$, $v$, and $i$ are the observed aperture photometry
magnitudes, $B$, $V$, and $I$ are the standard system magnitudes, and
$X$ is airmass. The transformation coefficients were determined using
the program CCDSTD (e.g. Stetson 1992).  While it was clear that higher 
order color terms would be necessary to adequately fit measurements of
extremely red stars ($\bv > 2.0$), we found that such terms were
unnecessary because the cluster stars fell in a range of colors that was
quite well fitted by linear color terms. 
Our calibrated measurements for the standard stars are compared to the catalog
values are shown in Figure~\ref{primary}.

\subsection{Calibration Against Secondary Standards}

Observations in each filter of the cluster fields were made on night 6, and
used to calibrate the cluster data. We selected 193 stars with 
relatively low measurement errors from the
outskirts of the cluster as our local standards. 
These stars were generally on the asymptotic giant branch, upper RGB, 
or horizontal branch (HB),
and covered the entire range of colors for the cluster stars
observed. We used the photometric transformations above to derive
standard values for these stars.

\begin{figure}
\plotone{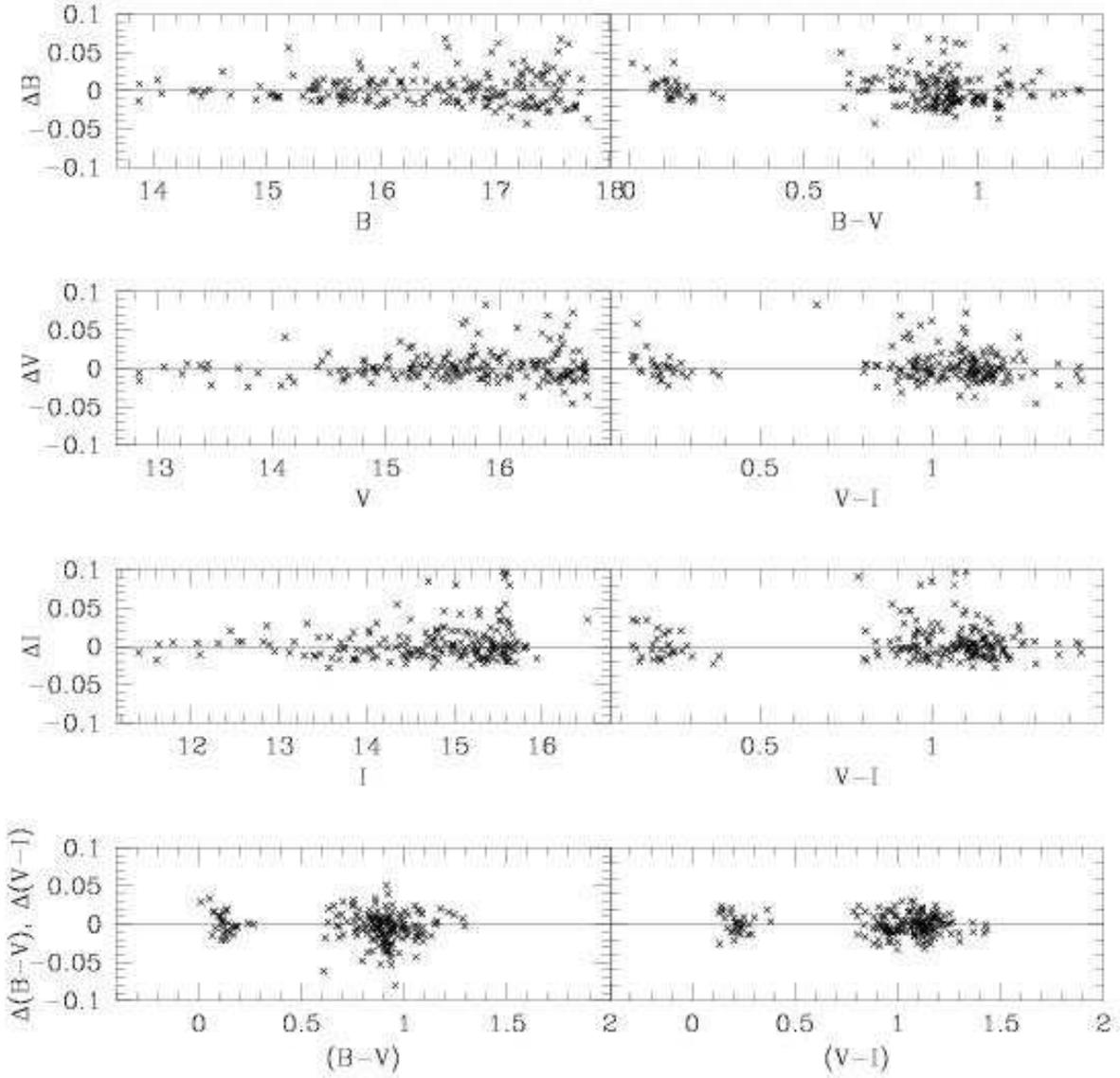}
\caption{Photometric residuals (in the sense of the final point-spread
function photometry minus standard aperture photometry values) from
the calibration of secondary standard stars.
\label{secondary}}
\end{figure}

The calibrated secondary standard values were then used to calibrate
the PSF-fitting photometry. PSF-fitting photometry from both nights of
M12 observations were combined and averaged after zero-point
differences between frames had been determined and taken into account.
We then verified that the linear color terms derived earlier
accurately corrected our data for color-dependent systematic errors
(see Figure~\ref{secondary}), and determined zero-point corrections
for the photometry in each filter band.  As a final note, we did not
include the measurements of the brightest calibrated stars from the
longest exposed $V$-band frames in order to avoid introducing systematic
errors from non-linearity near CCD saturation.

\subsection{Comparison to Previous Studies}

\begin{figure}
\plotone{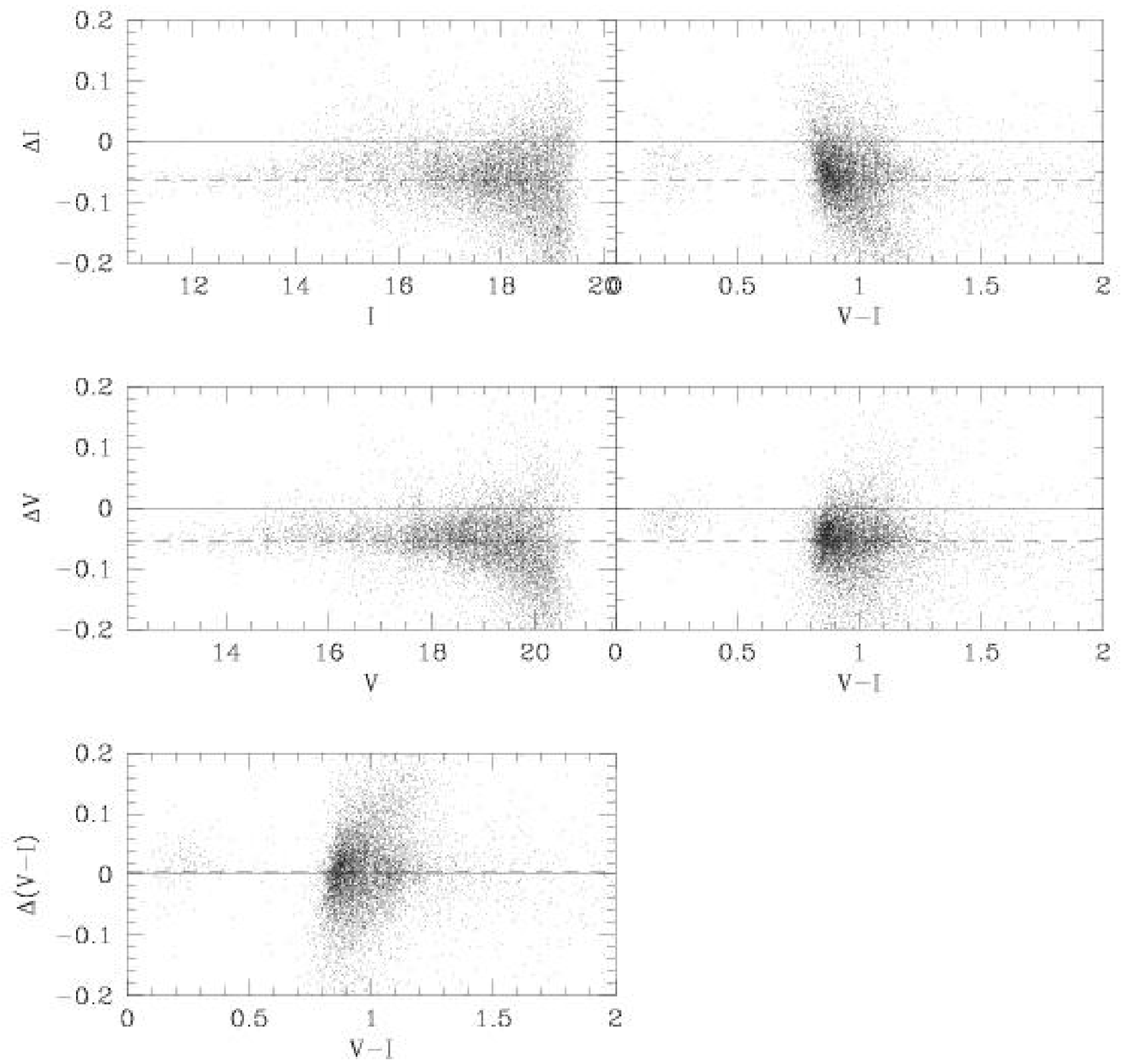}
\caption{Residuals (in the sense ours minus theirs) from the star-by-star
comparison of our photometry with that of \citet{vb02}.  The dashed line 
represents the median value of all points.
\label{compkasp}}
\end{figure}

\begin{figure}
\plotone{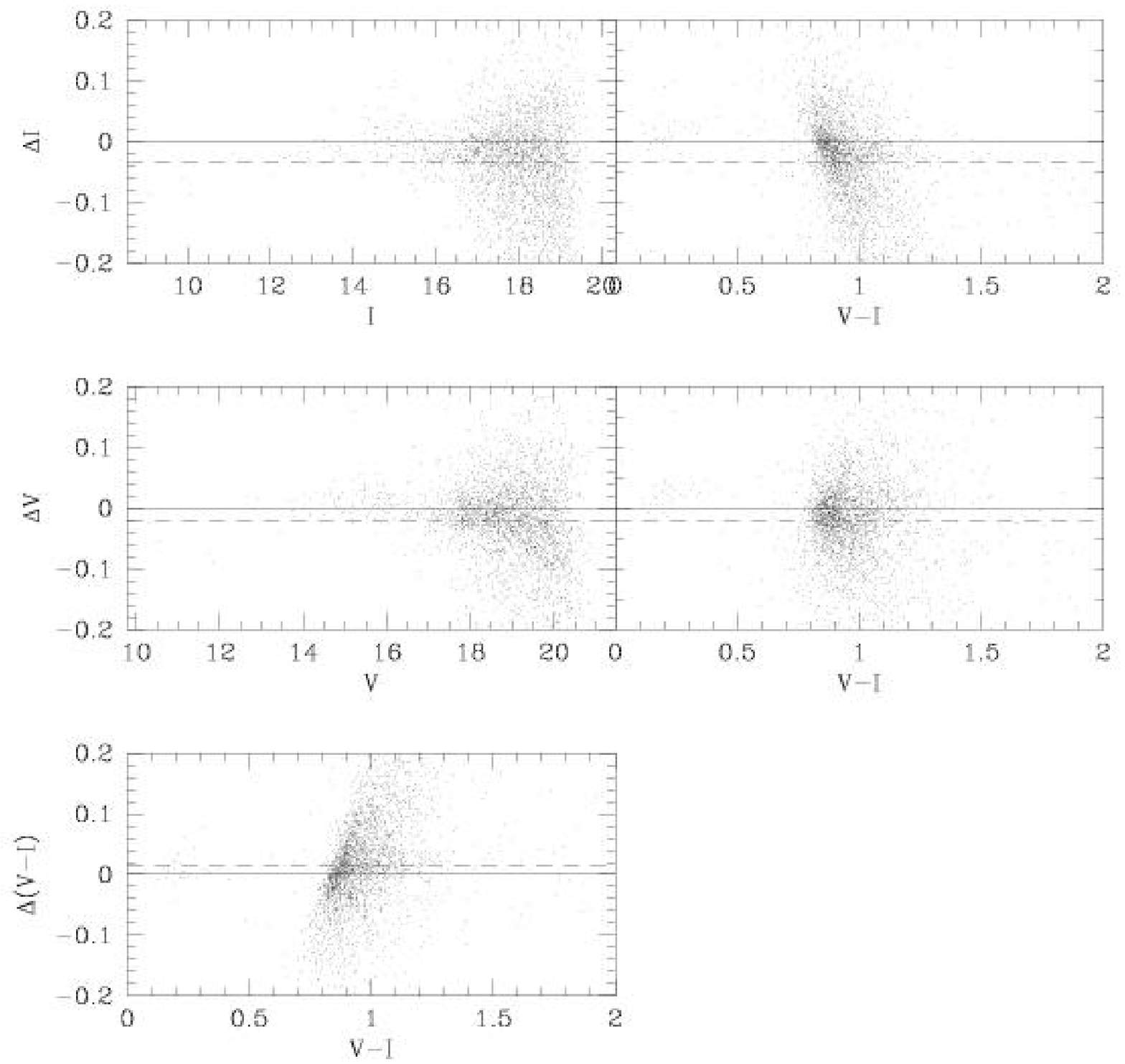}
\caption{Residuals (in the sense ours minus theirs) from the
star-by-star comparison of our photometry with that of \citet{ro00}.  The
dashed line represents the median value of all points.
\label{comprosen}}
\end{figure}

\begin{figure}
\plotone{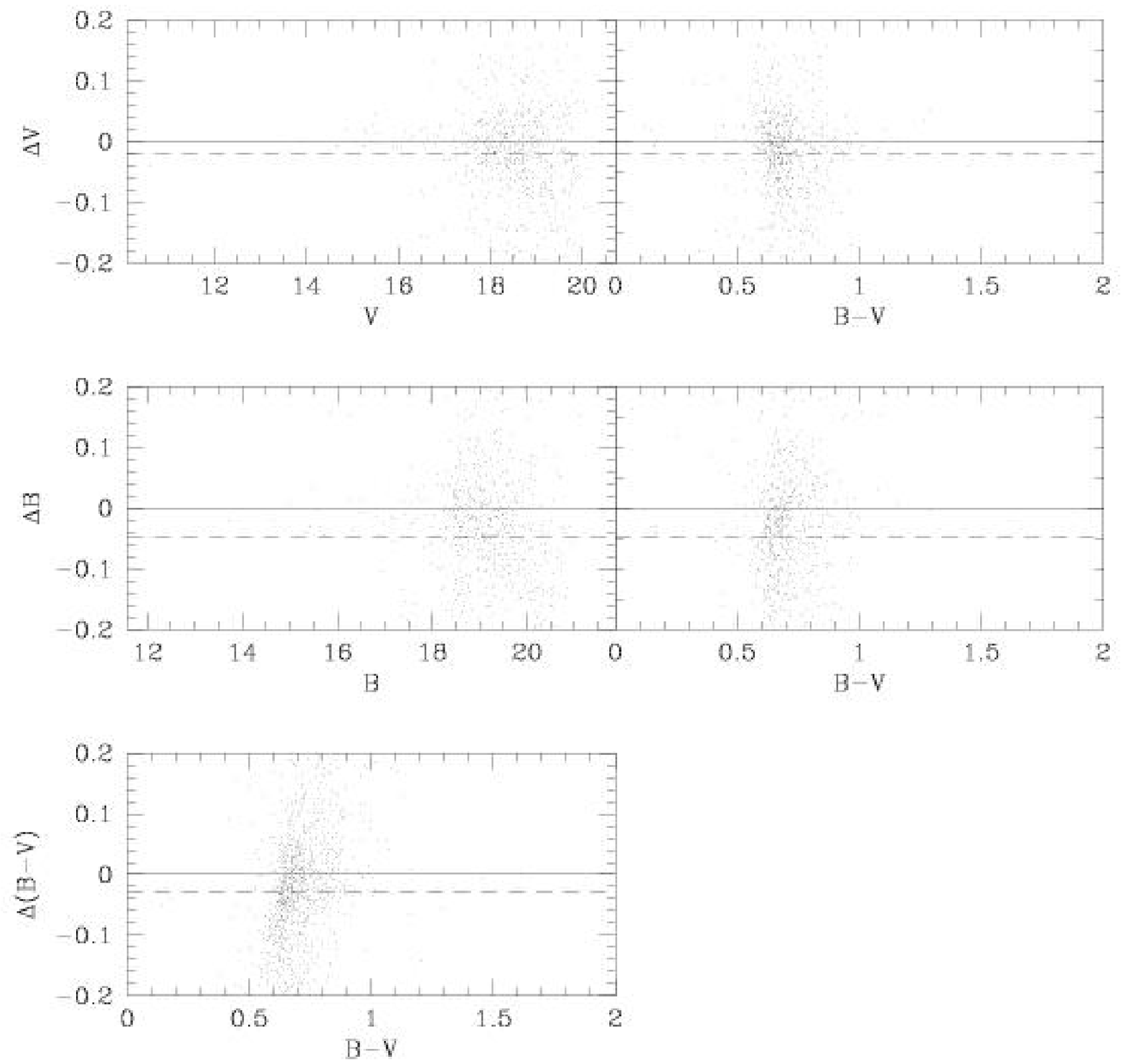}
\caption{Residuals (in the sense ours minus theirs) from the
star-by-star comparison of our photometry with that of \citet{br96}.  The
dashed line represents the median value of all points.
\label{compbrocato}}
\end{figure}

\begin{figure}
\plotone{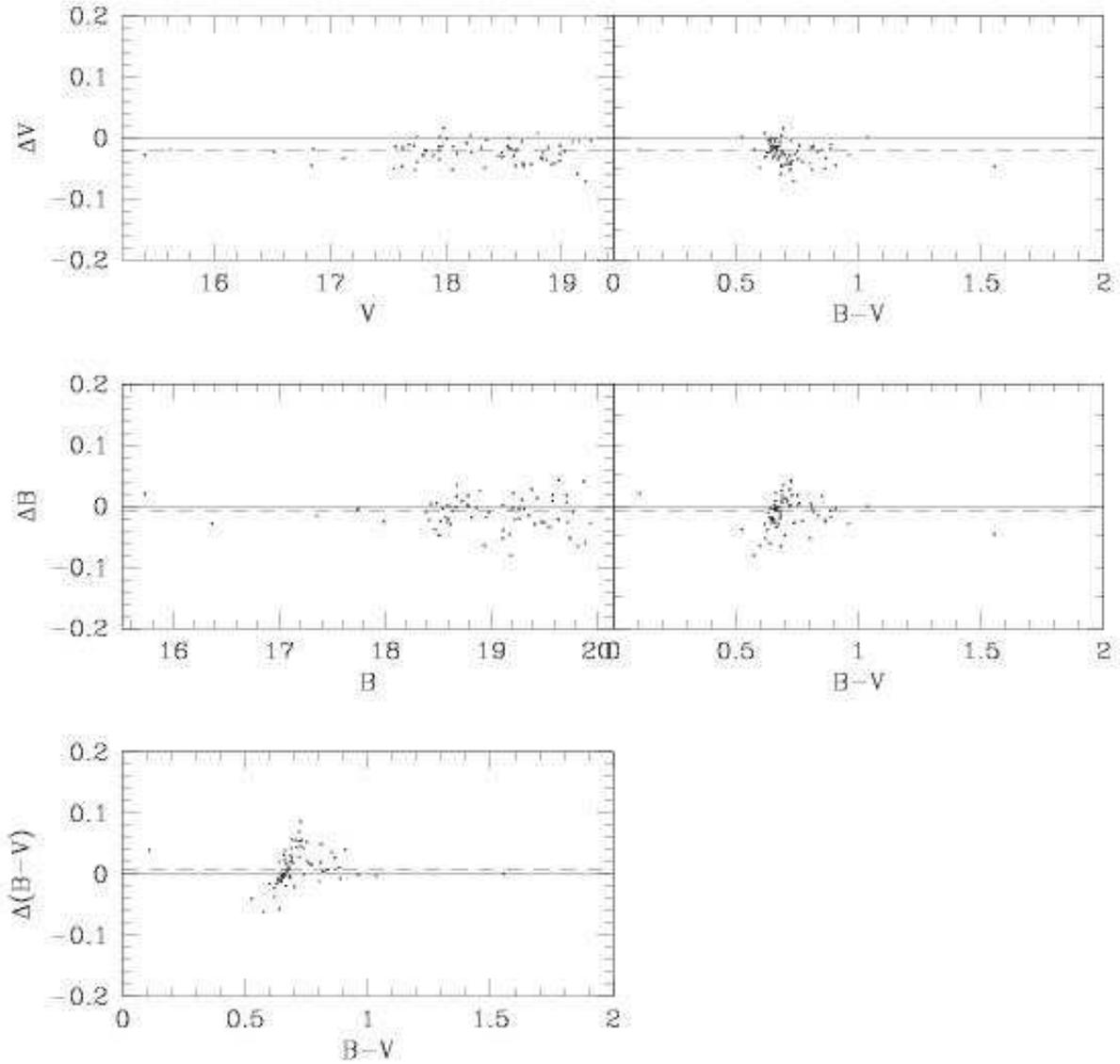}
\caption{Residuals (in the sense ours minus theirs) from the
star-by-star comparison of our photometry with the local standard stars 
of Stetson (2000) in this cluster.  The dashed line represents the median 
value of all points.
\label{compstet}}
\end{figure}

In order to check the accuracy of our photometric calibration, our
data set was compared (star-by-star) to recent ground-based data from
R00, VB02, and B96.  Figures~\ref{compkasp}
and~\ref{comprosen} show the $V,I$, and $(V-I)$ photometric residuals
(our data minus theirs) from comparisons with the VB02 and
R00 studies, respectively.  Figure~\ref{compbrocato} shows
the $V$ and $(B-V)$ residuals from comparison with the
B96 data.  In Table~\ref{comp} we provide the median values
of these residuals, since this statistic is less sensitive to ``outliers'' 
than the mean. Our data agree (within reasonable errors) with
both the B96 and R00 data.  The VB02 photometry is significantly faint 
compared to ours.  Because the median of the residuals ranges from $0.02$ 
to $0.06$ magnitudes, we also compare our data to the Stetson (2000; 
denoted S00 in Table~\ref{comp}) local standard stars in this cluster 
and show the residuals in Figure~\ref{compstet}.  This comparison yields
small median residuals showing consitent photometric calibration between
this study, the Stetson (2000) local standards, and the B96 \& R00 data
sets.

\section{The Color Magnitude Diagram}

\begin{figure}
\plotone{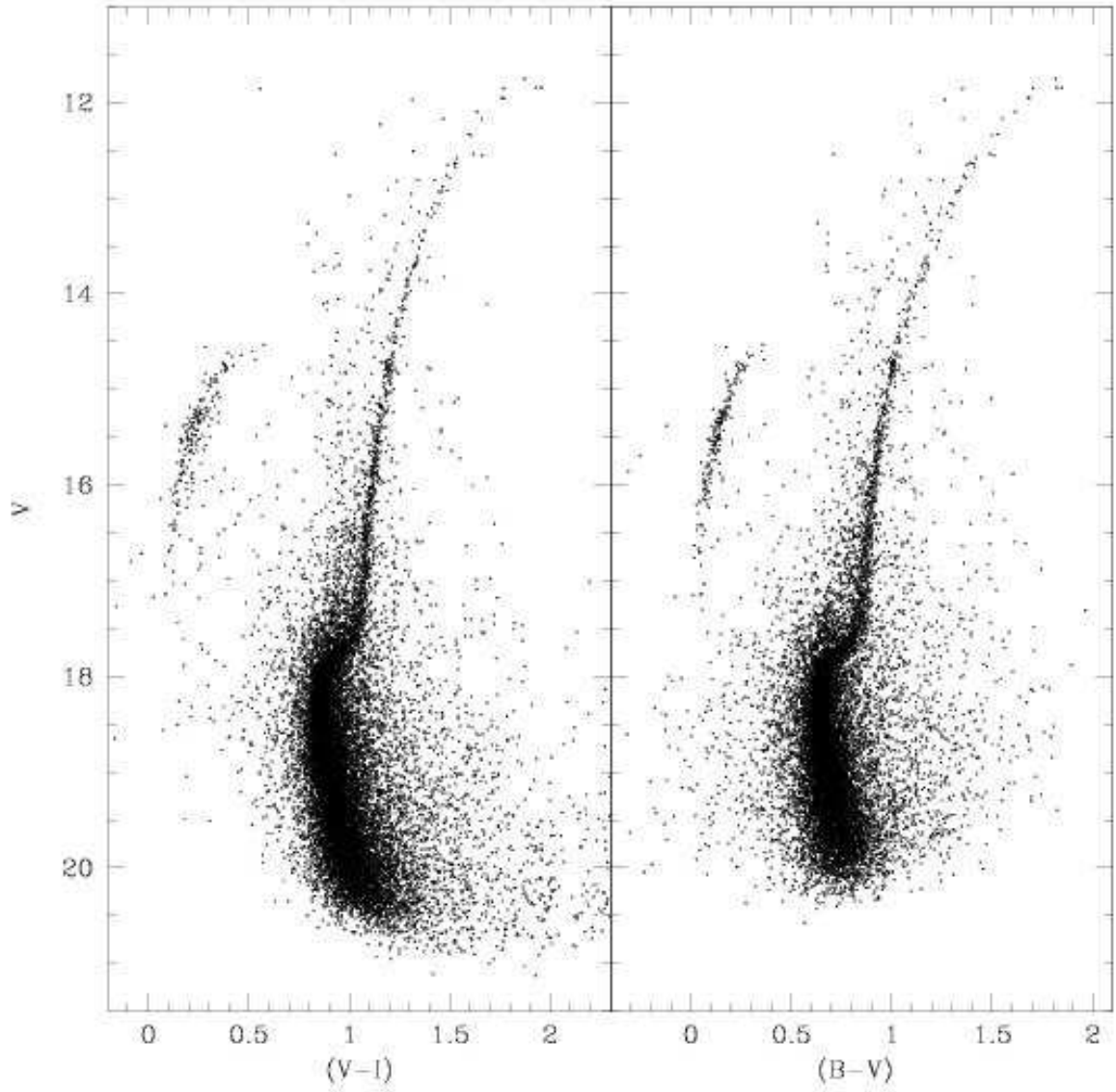}
\caption{$V,(V-I)$ and $V,(B-V)$ CMDs for all 17,303 stars measured
in this study.
\label{cmds}}
\end{figure}

\begin{figure}
\plotone{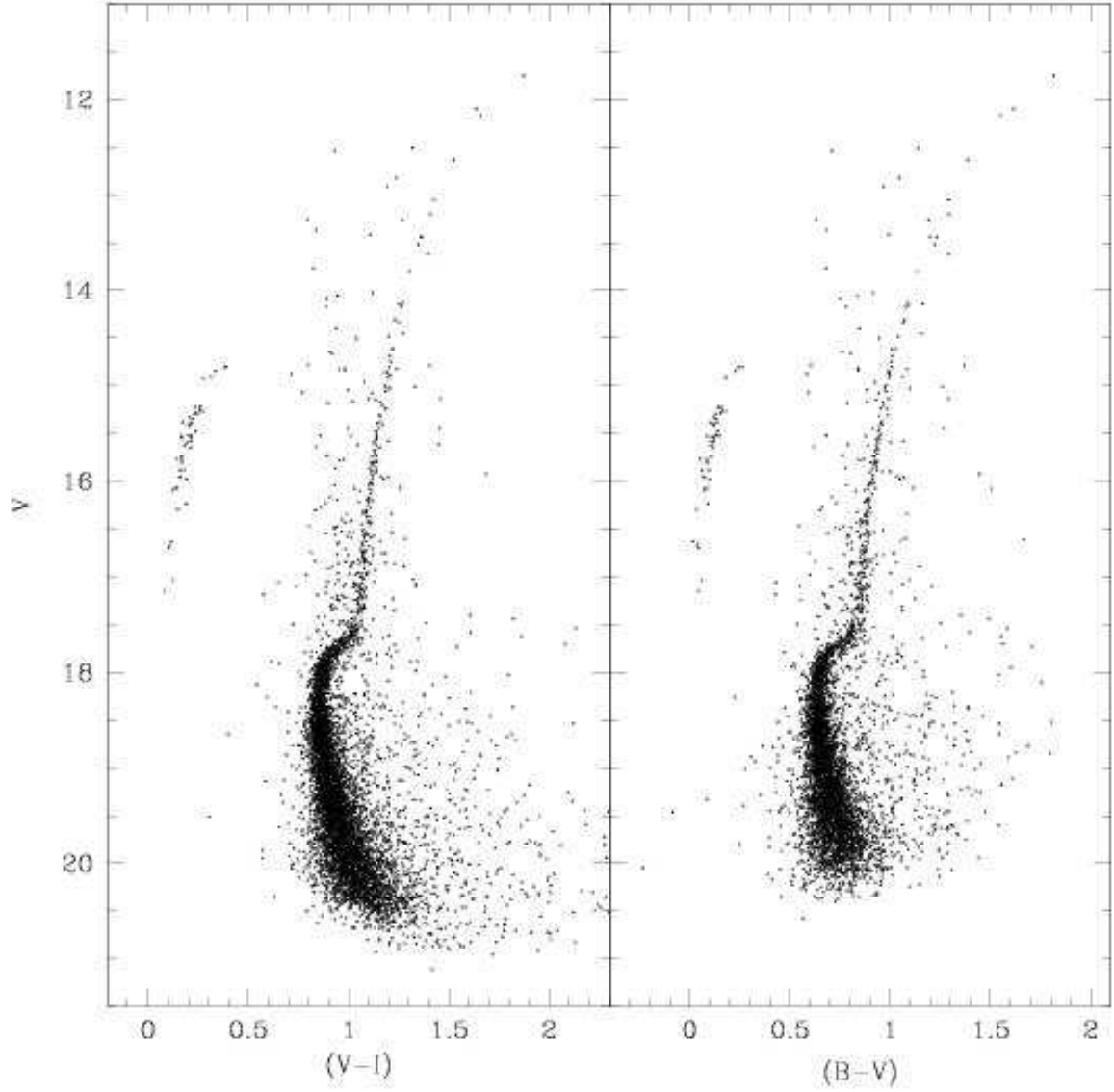}
\caption{$V,(V-I)$ and $V,(B-V)$ CMDs for the sample of measured
stars restricted to having a radius between 3\Min4 and 8\Min5 from the cluster
center.
\label{cmdcutrad}}
\end{figure}

The results of this $BVI$ photometric study are presented as CMDs in
Figures~\ref{cmds} and~\ref{cmdcutrad}.  The total sample of 17,303 stars 
measured in this study is shown in Figure~\ref{cmds}.  Given the lack of
structure in the CMD beyond 8\Min5, in our final sample we ignored stars
beyond this radius from the cluster center.  Figure~\ref{cmdcutrad} shows the
CMDs of the cluster restricted to those stars located between a radius of
3\Min4 and 8\Min5 from the cluster center. We derive 
fiducial sequences for both
the $V,(B-V)$ and $V,(V-I)$ CMDs, and present the data in
Tables~\ref{bvfid} and~\ref{vifid}, including the number of stars $N$ in
each bin used to compute the fiducial point.
For the MS, the fiducial sequence was determined by
taking the mode of the color distribution in magnitude bins. The
SGB fiducial points were also determined by finding
the mode of the magnitude distribution in color bins because of the
horizontal nature of the SGB in the CMD.  The mean of the color
distribution was used to compute fiducial points for the RGB. The mean of 
the distribution in a combination of color
and magnitude bins were used to compute the fiducial points for the
HB.

\begin{figure}
\plotone{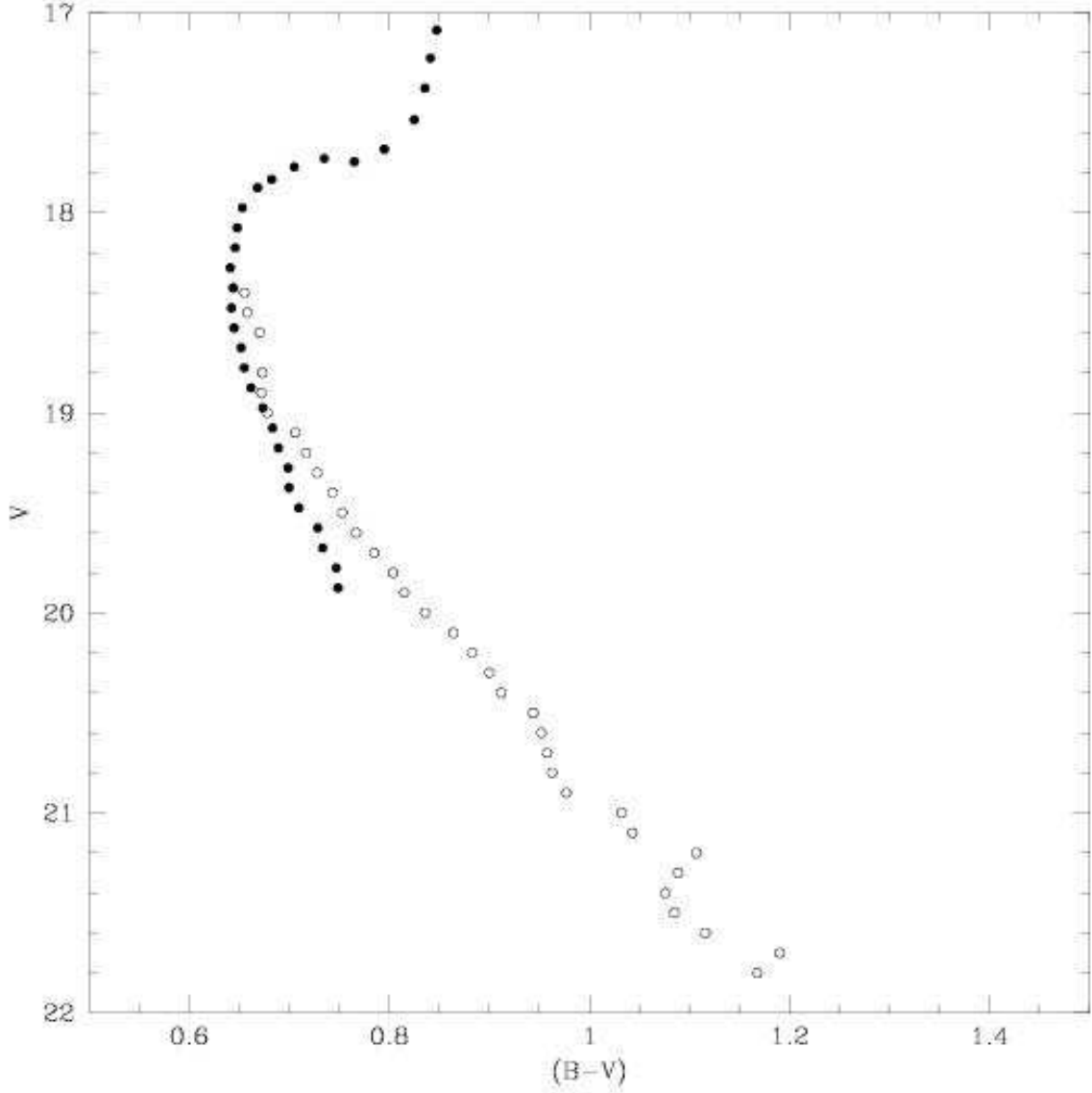}
\caption{Comparison of the fiducial sequence derived in this study
(\textit{filled circles}) with that of \citet{sa89} (\textit{open
circles}).
\label{satocomp}}
\end{figure}

In Figure ~\ref{satocomp}, we compare our derived $V,(B-V)$ fiducial
sequence with that of S89.  Their $UBV$ photometric study of
M12 presents the only recent fiducial sequence available for
comparison to our data set.  We attribute the differences in the slope
and offset of the MS fiducials to differences in the photometric
calibrations, although this is difficult to verify since no other
published study has done star-to-star comparisons with the S89 dataset.

\section{Cluster Parameters: Metallicity, Reddening, Distance Modulus, and 
Age}

In this section we describe our method for the determination of four
cluster parameters (metallicity, reddening, distance modulus, age) 
necessary to compare the theoretical LF to the observed.

\subsection{Metallicity}

There have been a number of [Fe/H] studies of M12, and published values
range over nearly $0.5$ dex.  The two most widely used
metallicity scales are those of Zinn \& West (Zinn \& West 1984; Zinn
1985; hereafter ZW) and Carretta \& Gratton (1997; hereafter CG).  ZW
cite a value of [Fe/H]$=-1.61$.  The CG scale (based on
high-resolution spectroscopy of GGC red giants) gives [Fe/H]$=-1.37$
from the quadratic transformation of the ZW scale. The discrepancy
between the two scales is well-documented, with the CG scale giving a
higher metallicity by approximately $0.2-0.3$ dex for low- or
intermediate-metallicity clusters (such as M12) and approximately
$0.1$ dex lower abundances for metal-rich clusters.  Spectroscopic
measurements of the infrared \ion{Ca}{2} triplet of M12 red giants
have been made by \citet{su93} and \citet{rhs97a}.  \citet{rhs97b} used
these measurements to compute abundances on
the ZW and CG scales of [Fe/H]$_{ZW}=-1.40\pm0.07$ and
[Fe/H]$_{CG}=-1.14\pm0.05$.  Recent work by \citet{kr03} finds a
metallicity of $\textrm{[Fe/H]}_{KI}=-1.25$ from observations of the
equivalent width of \ion{Fe}{2} in cluster red giants and calibration
with $W'$ from \citet{rhs97b}. For the remainder of this study, we
only consider metal abundances of M12 between
$-1.40<\textrm{[Fe/H]}<-1.14$\footnote{A metallicity of
[Fe/H]$_{ZW}=-1.61$, however, is used for some comparisons of our
observations to the theoretical luminosity functions of \citet{bv01}.  
See $\S6.1$ and $\S6.2$ for more details}.

\subsection{Reddening}

In this study we adopt the reddening values as determined by VB02.
They note the lack of significant differential reddening across the
field of M12, and hence we do not use their maps to internally
deredden our data.  We use their mean reddening value of
$E(V-I)=0.25$ that is in agreement with the infrared dust emissivity
maps of \citet{sc98} who also find $E(V-I)=0.25$.  Other measured values for
the cluster reddening range from $E(B-V)=0.17-0.23$ (Racine 1971; S89).  Given 
the agreement between the VB02 and Schlegel et al. (1998) studies, we adopt a 
value of $0.02$ as the uncertainty in the reddening.

\subsection{Distance Modulus}

Previous determinations of the distance modulus $(m-M)_V$ of M12 have
yielded a wide range of values, from $(m-M)_V=14.02$ (VB02) to 14.30
\citep{ra71}. Even between studies that adopt similar techniques to find the
distance modulus (namely subdwarf fitting) the results are not in agreement:
the study by S89 finds $(m-M)_V=14.25\pm0.20$ but \citet{so01} find
$(m-M)_V=14.03\pm0.11$. Given that the overall uncertainty in previous
distance determinations is inadequate to define a well-constrained range, we 
use the technique of subdwarf fitting to re-determine the
distance modulus of M12. 
Because the data in this study are mostly drawn from the evolved stellar
populations, our MS is not
faint enough to be adequate for this fitting technique. To overcome this, we
use the VB02 data which goes several magnitudes fainter and has a well-defined
MS. Fiducial points (listed in Table~\ref{vbfid}) for the main
sequence were determined using methods identical to those described in $\S3$,
after correcting the data for the median offsets in Table~\ref{comp}.
In order to minimize the possibility of systematic effects in the
distance determination, we adopt the CG metallicity scale for both the
subdwarfs and M12.  We limit our sample of possible
subdwarfs to those that have well-determined parallaxes $\pi$
(specifically those with relative error $\sigma_\pi/\pi<0.12$).  This
list was further restricted to stars that have metallicities measured
on the CG scale.  This subset was further limited by lack of $I$-band
photometry: measured $V-I$ colors are sparse for metal-poor subdwarfs
in the literature.  From these considerations, our list of subdwarfs
has magnitudes in the range $4.56<M_V<7.17$, metallicities between
$-1.79<$[Fe/H]$<-0.90$, and relative parallax errors $\sigma_\pi/\pi<0.08$.

These 13 potentially usable subdwarfs are listed in Table
\ref{sdlistvi}.  Columns 1 and 2 list the \textit{Hipparcos} Input
Catalog number and HD (or Gliese) number, respectively. Columns 3 and
4 list the reddening $E(B-V)$ and apparent $V$ magnitude, as compiled
by \citet{ca00} from the photometry of \citet{ca94,ry91,sc89} and the
\textit{Hipparcos} catalog.  Columns 5 and 6 give the
\textit{Hipparcos} parallax $\pi$ (in units of milliarcseconds) and
the relative parallax error $\sigma_\pi/\pi$, both taken from the
catalog.  The absolute $V$ magnitude $M_V$ is listed in column 7 (and
its error $\sigma_{M_V}$ in column 8) and includes the Lutz-Kelker
corrections following the procedure described by \citet{ha79}.  The
observed $(V-I)$ colors in column 9 are taken from \citet{de81},
\citet{ma96}, and \citet{re01}.  The metal abundance [Fe/H] on the CG
scale from \citet{ca00} is listed in column 10.  The deviation between
the observed subdwarf color and the theoretically predicted color,
denoted as $\delta(V-I)$, is listed in column 11.  Column 12 shows the
subdwarf colors after application of the theoretical color correction.
For our MS fit, we use only those subdwarfs having metal
abundances in the range $-1.50<\textrm{[Fe/H]}<-1.20$, following the
discussion in \citet{va00,va02}.  They note the excellent agreement of
the observed and theoretical colors for the subdwarfs in this range.
Our results confirm this agreement; the mean deviation of the observed
and theoretical colors is $0.004$ for the six subdwarfs in this range.
Our final list of 6 subdwarfs used in the fit have absolute magnitudes
$M_V>5$, metallicities between $-1.48<$[Fe/H]$<-1.24$ (mean of $-1.35$),
and relative parallax errors $\sigma_\pi/\pi<0.070$.  We emphasize
that this analysis assumes an $\alpha$-element abundance enhancement
of $[\alpha$/Fe]$=+0.3$ for each subdwarf and negligible age differences.

\begin{figure}
\plotone{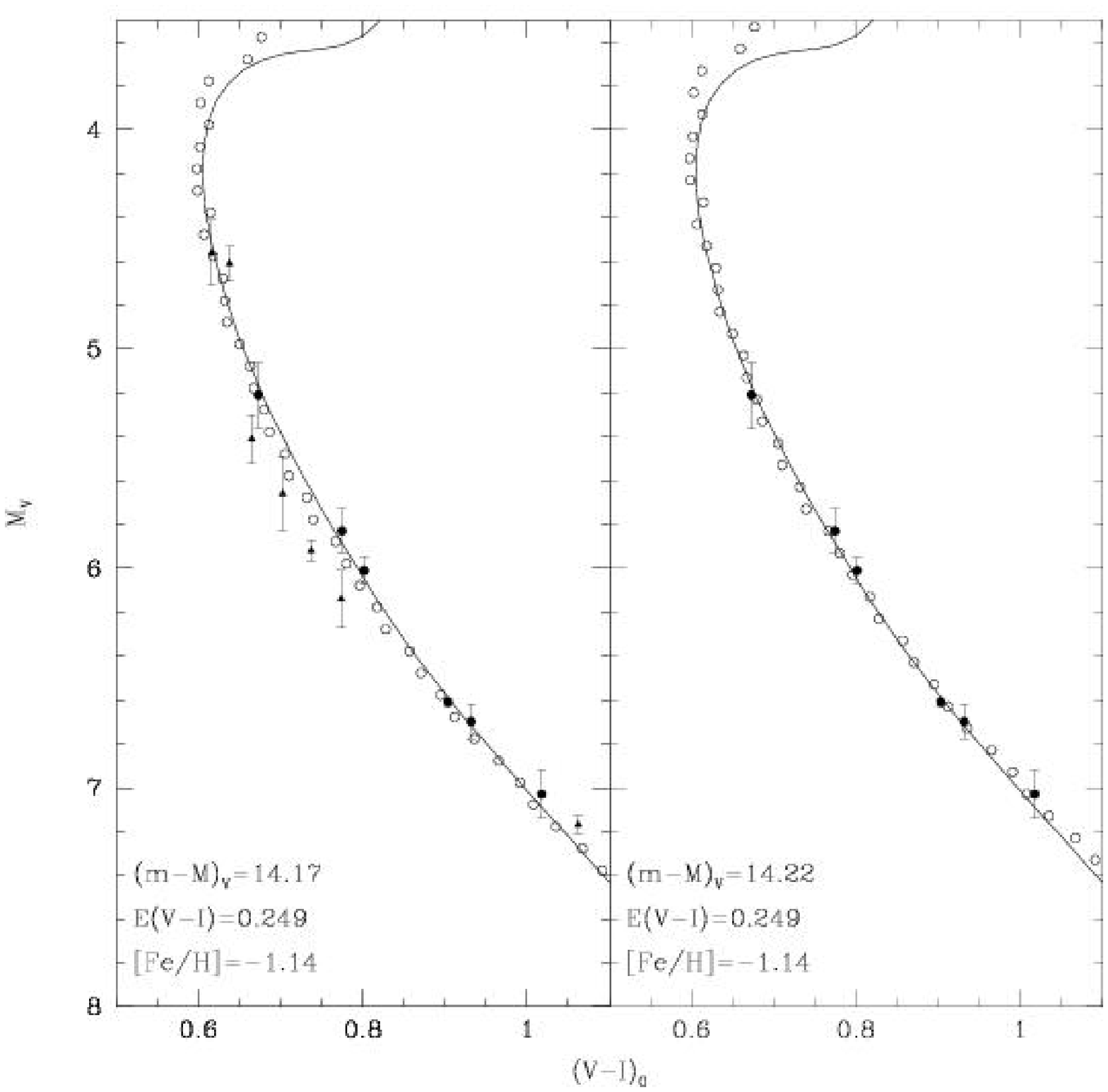}
\caption{Example of the subdwarf fitting performed on the MS fiducial of
\citet{vb02} (\textit{open circles}).  The fiducial points have
been shifted to the derived value of $(m-M)_V$.  Overlaid on the
subdwarfs (\textit{filled circles}) is the 12 Gyr isochrone from 
\citet{bv01} (\textit{solid line}) for a metallicity of [Fe/H]$=-1.14$.  The
left panel shows the fit using all 13 subdwarfs; the right panel shows the
fit using the restricted set of 6 subdwarfs.  Those subdwarfs denoted as
triangles were eliminated from the final distance modulus determinations.
\label{dm}}
\end{figure}

\begin{figure}
\plotone{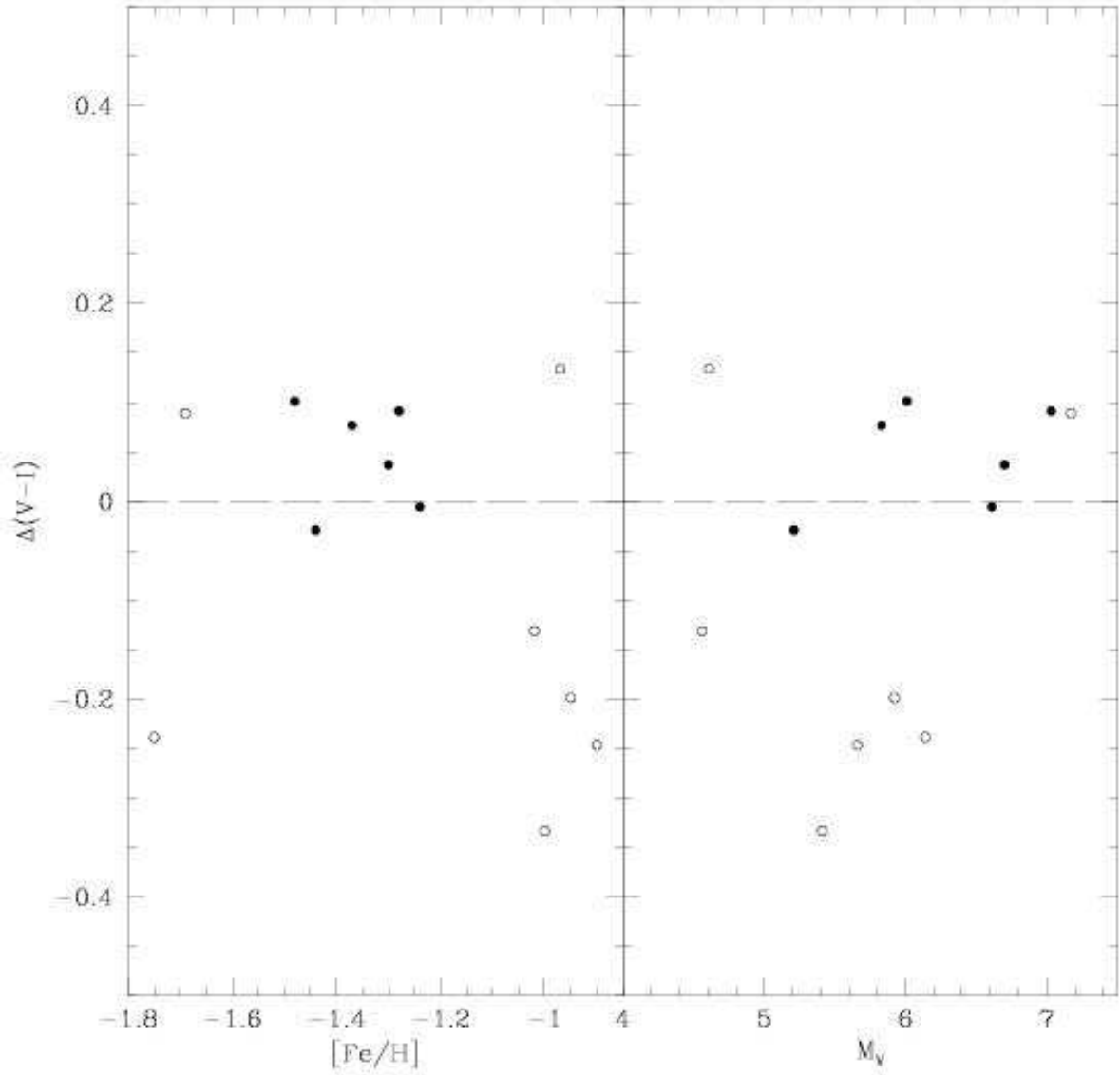}
\caption{Residuals to the distance modulus fit in Figure~\ref{dm} as a
function of metallicity and absolute $V$ magnitude.  The stars denoted as open circles
(which correspond to the triangles in Figure~\ref{dm}) were eliminated from the final
distance modulus determinations.
\label{dmres}}
\end{figure}

Figure~\ref{dm} shows the best fit of the M12 fiducial to these stars,
along with a 12 Gyr isochrone from Bergbusch \& VandenBerg (2001;
hereafter BV) for $\textrm{[Fe/H]}=-1.14$.  The derived distance
modulus changes depending on which set of subdwarfs are selected
[$(m-M)_V=14.17$ for all 13 subdwarfs; $(m-M)_V=14.22$ for our final
list of 6 subdwarfs]. We used a polynomial interpolation between
several 12 Gyr isochrones of BV to determine a theoretical color
correction for each subdwarf.  This correction is computed as the
difference at the $M_{V}$ of the subdwarf between the colors of
isochrones having the metallicity of M12 and the metallicity of the
subdwarf.  In order to fit for the distance modulus, the fiducial of
M12 is shifted in magnitude to match each subdwarf individually.
Thus, each subdwarf provides a measure of the distance modulus and our
final estimate is a mean value weighted by the squares of the error
estimates of the absolute magnitude.  These error estimates include
the uncertainties in the subdwarf's parallax, reddening, and
metallicity.  We assume an uncertainty in the metallicity of each
subdwarf of 0.1 dex.  The derived apparent distance modulus of M12 (assuming 
a metallicity of [Fe/H]$=-1.14$) is $(m-M)_V=14.22\pm0.11$ using the six
subdwarfs in Table~\ref{sdlistvi}.  We plot in Figure~\ref{dmres} the
difference between the theoretically corrected subdwarf color and the M12
fiducial color (at the absolute $V$ magnitude of the subdwarf),
denoted as $\Delta(V-I)$, as a function of metallicity and absolute
$V$ magnitude.  These show no significant systematic errors from the
fit.  The largest uncertainty in the distance modulus comes from the
adopted cluster metallicity.  For metallicities [Fe/H]$=-1.41$ and
$-1.61$, we find distance moduli of $(m-M)_V=14.05\pm0.12$ and
$13.96\pm0.11$, respectively.

In order to check for possible systematic errors in the subdwarf color
corrections, we perform the same procedure of subdwarf fitting using
the Yonsei-Yale isochrones from Kim et al. (2002; hereafter Y$^2$) to
obtain the theoretical color correction to the subdwarfs.  We use the
color transformation table of \citet{gr87} (hereafter G87) to avoid
introducing any systematic errors from use of the \citet{le98}
(hereafter L98) table, which clearly differs from both the
BV color transformation and the G87 table at faint absolute
magnitudes. Using the Y$^2$ isochrones with the G87 tables, we find an
apparent distance moduli of $(m-M)_V=14.23\pm0.11, 14.05\pm0.12$ and
$13.94\pm0.12$ for metallicities of [Fe/H]$-1.14,-1.41,$ and $-1.61$.
These are in excellent agreement (within the errors) to the value derived from
the BV isochrones.

\subsection{Age}

The latest studies of the cosmic background radiation data from WMAP 
have found the age of the universe
to be $13.7\pm0.2$ Gyr \citep{sp03}, setting a tight upper-limit on
the possible ages of GGCs.  Using the \textit{relative} age indicator
$\Delta V_{\textrm{\textsc{to}}}^{\textrm{\textsc{hb}}}$ (defined as
the difference between the $V$ magnitude of the ZAHB and MSTO points),
the \citet{ro99} study (which uses the R00 homogeneous data
set) deduces a value of $\Delta
V_{\textrm{\textsc{to}}}^{\textrm{\textsc{hb}}}=3.60\pm0.12$ for M12.  They
find M12 to be coeval (within the errors) with the oldest clusters that have
metal abundances [Fe/H]$<-1.2$.  \citet{sa02} use both relative and
absolute age dating (with the R00 data set) and find ages of
$12.5\pm1.3$ Gyr or $12.7\pm1.3$ Gyr for metallicities of
$\textrm{[Fe/H]}_{CG}=-1.14$ or $\textrm{[Fe/H]}_{ZW}=-1.40$,
respectively, for M12.  Both \citet{ro99} and \citet{sa02} find an age
dispersion for clusters of intermediate metallicities (possibly as
high as 25\%) but the study by \citet{vb00} finds this dispersion to
be smaller.  Assuming the age of M12 to be coeval (or nearly coeval)
with these oldest clusters and allowing for a possible age dispersion,
we consider the range of possible ages of M12 to be between 11 and
13 Gyr. We discuss age further in \S 6.

\section{Determination of the Luminosity Function}

\subsection{Artificial Star Tests}

In order to properly determine an accurate LF a
calculation of incompleteness corrections must be made.  To quantify 
the incompleteness as a function of both
magnitude (corrections for faintness) and radius (corrections for
crowding), extensive artificial star tests have been performed.  We
mostly follow the prescription given by \citet{sand96} for the
calculation of incompleteness corrections and here simply present a
review of the methodology as it applies to our data set.

The artificial star tests were restricted to the $V$ and $I$ frames.
A theoretical LF was used to set the distribution of
artificial stars as a function of magnitude.  The fiducial line gives
the corresponding $I$ magnitude for an input $V$ magnitude from the
theoretical LF.  Artificial star magnitudes were
chosen to create a sufficient number of bright stars, while weighting
the distribution to the faint end of the CMD. Positions for the
artificial stars are chosen at random within a grid such that no
artificial stars can overlap (separations are no closer than
$2\times($PSF radius$)+1$ pixels).  The central portion of the grid is twice
as dense as the outer portion, and hence will place a higher
percentage of artificial stars in the most crowded regions of the
cluster. The grid itself was randomly shifted by a fraction of a bin
width from run to run. The ADDSTAR routine from DAOPHOT was used to
add properly-scaled PSFs to the frames. Approximately 2,100 stars were 
added per frame in an individual artificial star run.  The frames with
artificial stars are reduced in a manner identical to our initial
photometric procedures, and were compared to a control run that had no
artificial stars.  We conducted 39 artificial star runs that resulted
in total of 84,400 stars being placed and reduced.

\begin{figure}
\plotone{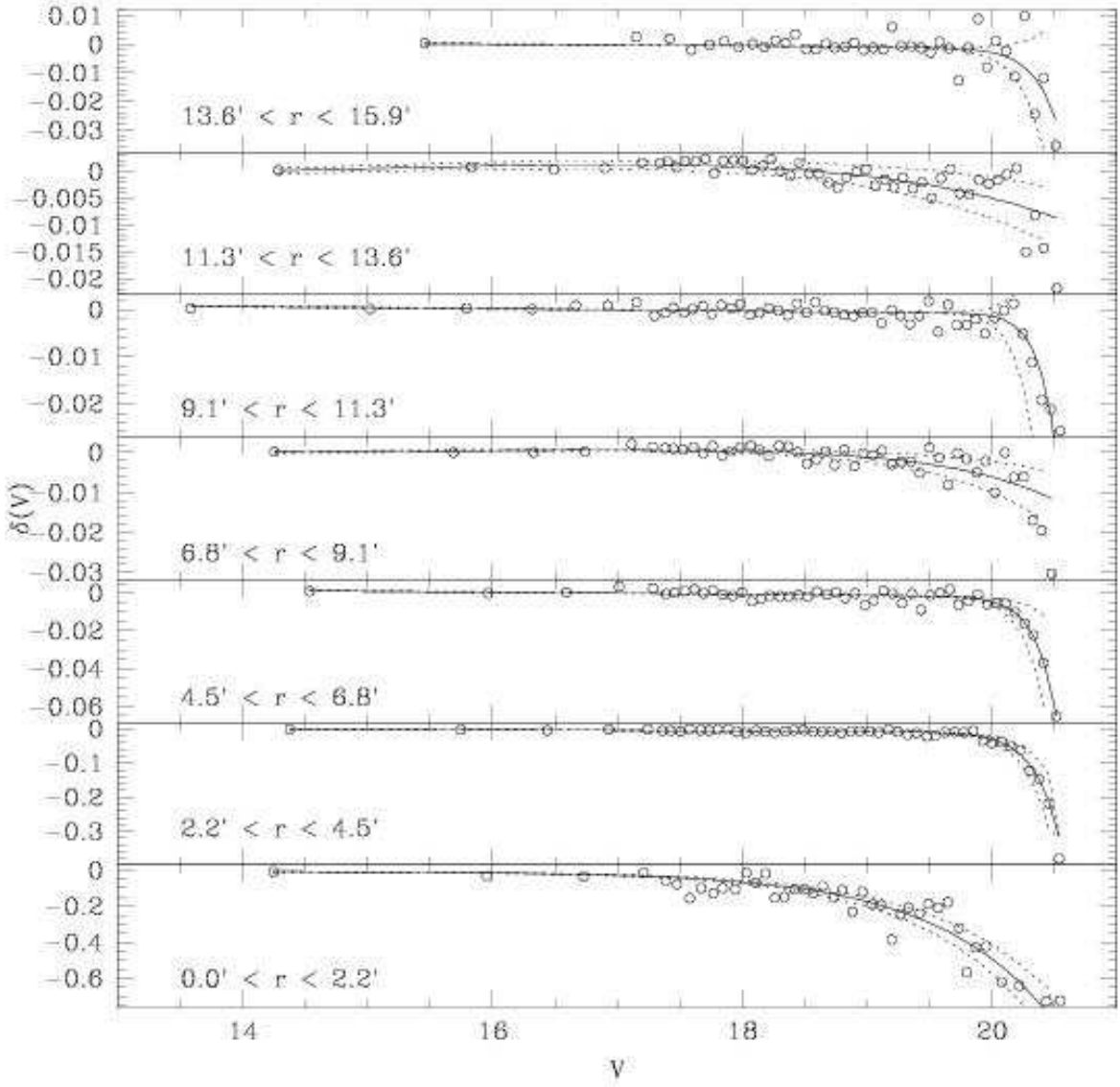}
\caption{Results from the artificial star tests for the
magnitude bias in the $V$ band $\delta(V)$ as a
function of radius and magnitude.
\label{deltav}}
\end{figure}

\begin{figure}
\plotone{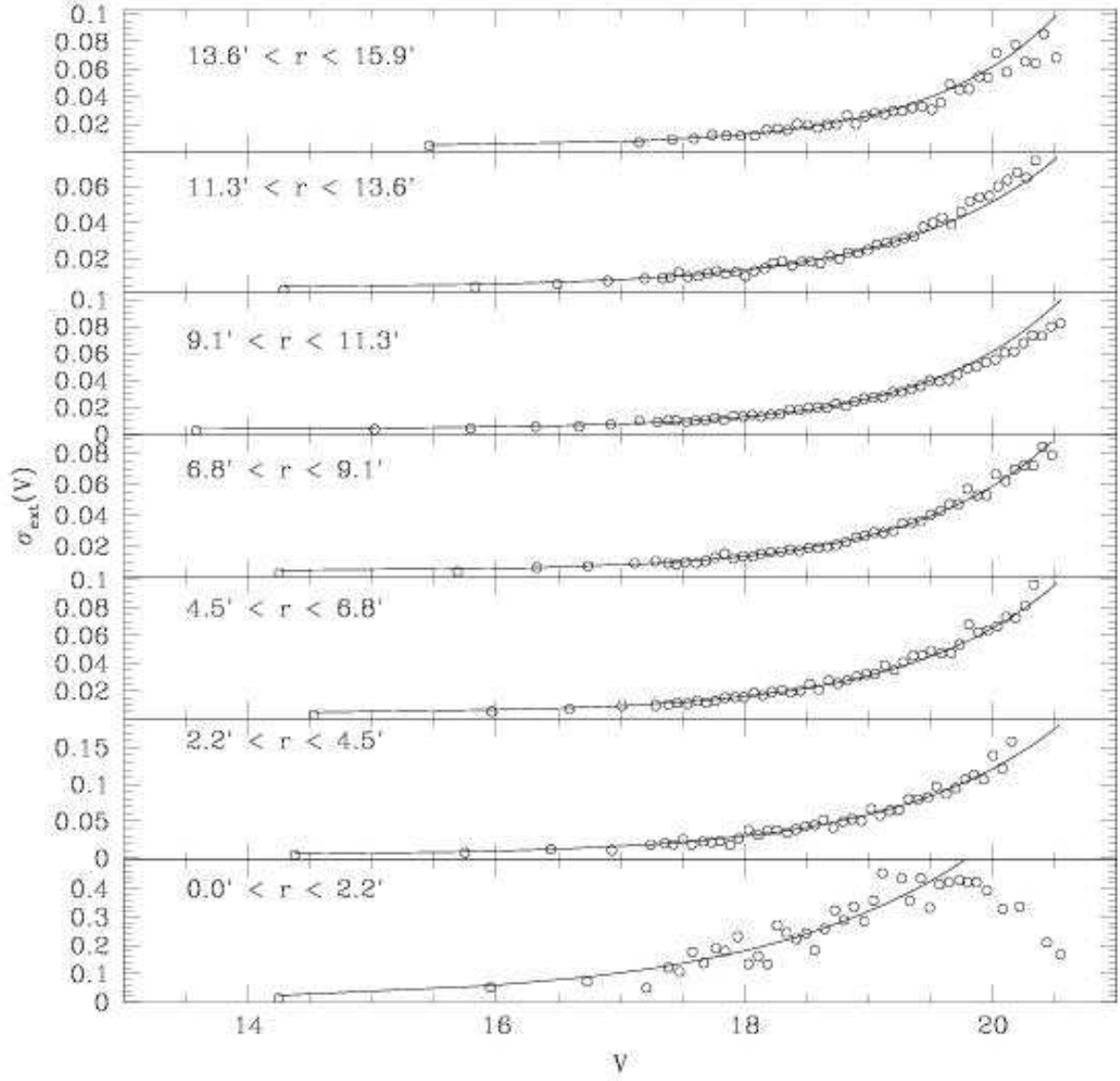}
\caption{Results from the artificial star tests for the external
$V$ magnitude errors $\sigma_{ext}(V)$ as a function
of radius and magnitude.
\label{sigmaev}}
\end{figure}

The output from the artificial star runs is a list of positions and
magnitudes for all detected stars.  What qualifies as a detection is
non-trivial; blending and crowding of artificial stars with real stars
will tend to favor the detection of the brightest stars (in a simple
positional search) regardless of whether or not they were artificial
(see \citet{sand96} for more details).  The resulting list of
recovered artificial stars is used to calculate the following
quantities (in bins sorted by projected radius and magnitude): (1)
median $V$ or $I$ magnitude, (2) median color $(V-I)$, (3) median
internal error estimates ($\sigma_V,\sigma_I,\sigma_{(V-I)}$), (4)
median magnitude and color biases ($\delta_V \equiv
\textrm{median}~(V_{\textrm{\small{output}}}-V_{\textrm{\small{input}}}),
\delta_I,\delta_{(V-I)}$), (5) median external error estimates
($\sigma_{\textrm{\small{ext}}}(V) \equiv
\textrm{median}~|\delta_V-\textrm{median}~(\delta_V)|/0.6745,
\sigma_{\textrm{\small{ext}}}(I),\sigma_{\textrm{\small{ext}}}(V-I)$),
and (6) total recovery probabilities ($F(V), F(I)$; the fraction of
stars added that were recovered at any magnitude).  In order to obtain
an estimate of these quantities beyond the magnitude limit of the
tests, we fit these quantities with the functional forms given in
\citet{sand96} and computed errors following the procedure in
\citet{sa99}. Figures~\ref{deltav}-\ref{sigmaev} present the results
of the above calculations for 200 pixel ($2\farcm3$) radial bins in
both $V$ and $I$ bandpasses.

\subsection{The Observed Luminosity Function}

\begin{figure}
\plotone{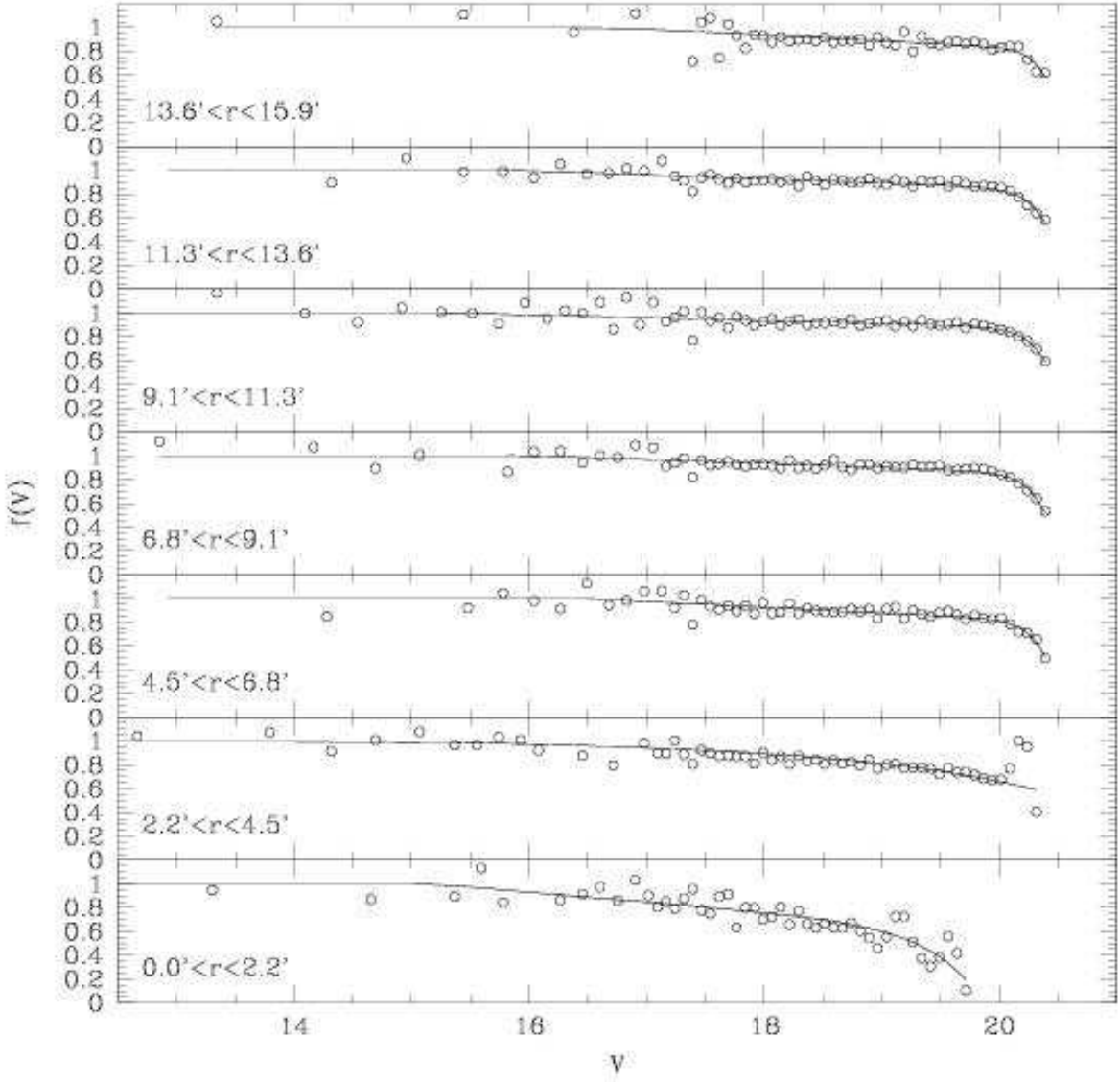}
\caption{Results from the artificial star tests for the completeness
fraction in the $V$ band as a function of magnitude and radius.
\label{fv}}
\end{figure}

\begin{figure}
\plotone{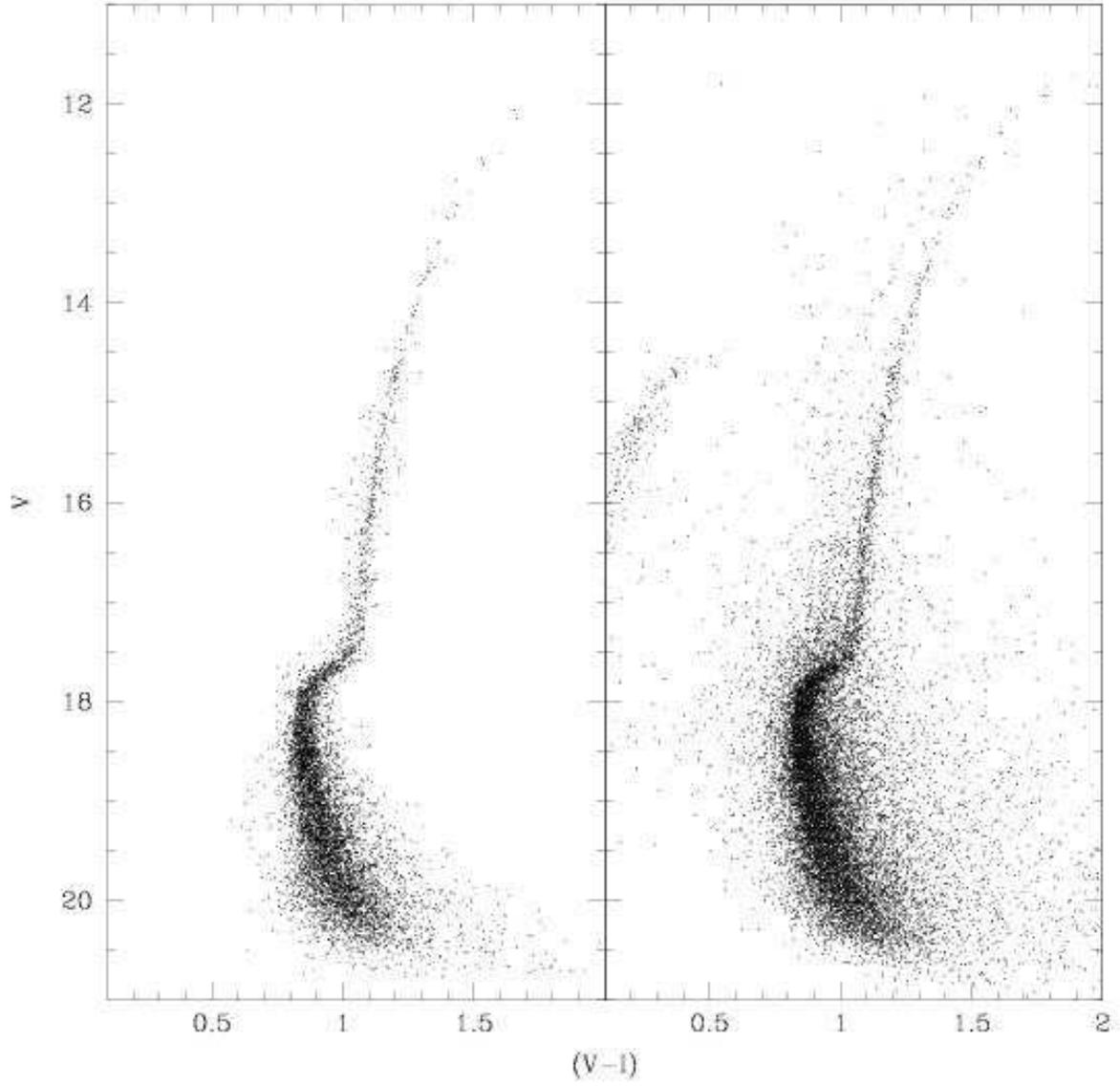}
\caption{Comparison of the stars kept (\textit{left panel}) out of the
total sample of star (\textit{right panel}) for the artificial star
tests.
\label{kept2}}
\end{figure}

From the results of the artificial star tests we determined the
corrections to the observed LF, following the
procedure of \citet{sand96} that is based on work of \citet{be93},
\citet{st88}, and \citet{lu74}.  In the computation of the LF the
error distributions, magnitude bias, and recovery probability are used to
predict the form of the observed LF when given an initial
estimate of the ``true'' LF.  Once the true
LF is determined, the completeness correction $f$ can
be calculated as simply the ratio of the predicted number of observed
stars to the actual number of observed stars.  The values of $f$ for
the various radial bins were fit using the same functional form as
$F$, and are plotted in Figure~\ref{fv}.  The total LF
was calculated using the completeness factor (multiplying each star by
the value $f^{-1}$ corresponding to its projected radius) and binned.
In Tables~\ref{volf} and~\ref{iolf} we present the observed $VI$ band
LF derived in this study, including the upper and lower $1\sigma$
error bars ($\sigma_{high}$ and $\sigma_{low}$, respectively).  
Figure~\ref{kept2} shows the CMD of those stars kept for the determination 
of the LF compared to the original sample.

\section{Comparison to Theoretical Models}

\subsection{The Color Magnitude Diagram}

Using the parameters derived in $\S4$, we compare our data to the BV
and Y$^2$ theoretical isochrones. For the Y$^2$ models (Version 2), we
compared our data to the isochrones computed using both the G87 and
L98 $T_{\mbox{eff}}$-color transformation tables.  Also included in this
comparison is the intermediate metallicity isochrone from \citet{va02}
(from the models of \citet{ri02} and \citet{tu98}; hereafter denoted
as the Richard \& VandenBerg model\footnote{\citet{va02} only
generated isochrones for two metal abundances, namely [Fe/H]$=-2.31$ and
$-1.31$.  We adopt the latter isochrones, which is in the range of the
metallicity of M12 determined in $\S4.1$}). The BV and Richard \& 
VandenBerg isochrones are computed from identical $T_{\mbox{eff}}$-color 
transformation relations which are a preliminary version of those 
presented by VandenBerg \& Clem (2003; D. A. VandenBerg 2004, private 
communication). All isochrones have been computed assuming 
an $\alpha$-element abundance enhancement of $[\alpha$/Fe]$=+0.3$.  
The differences between the input physics of the models are noted in 
Table~\ref{physics}.

\begin{figure}
\plotone{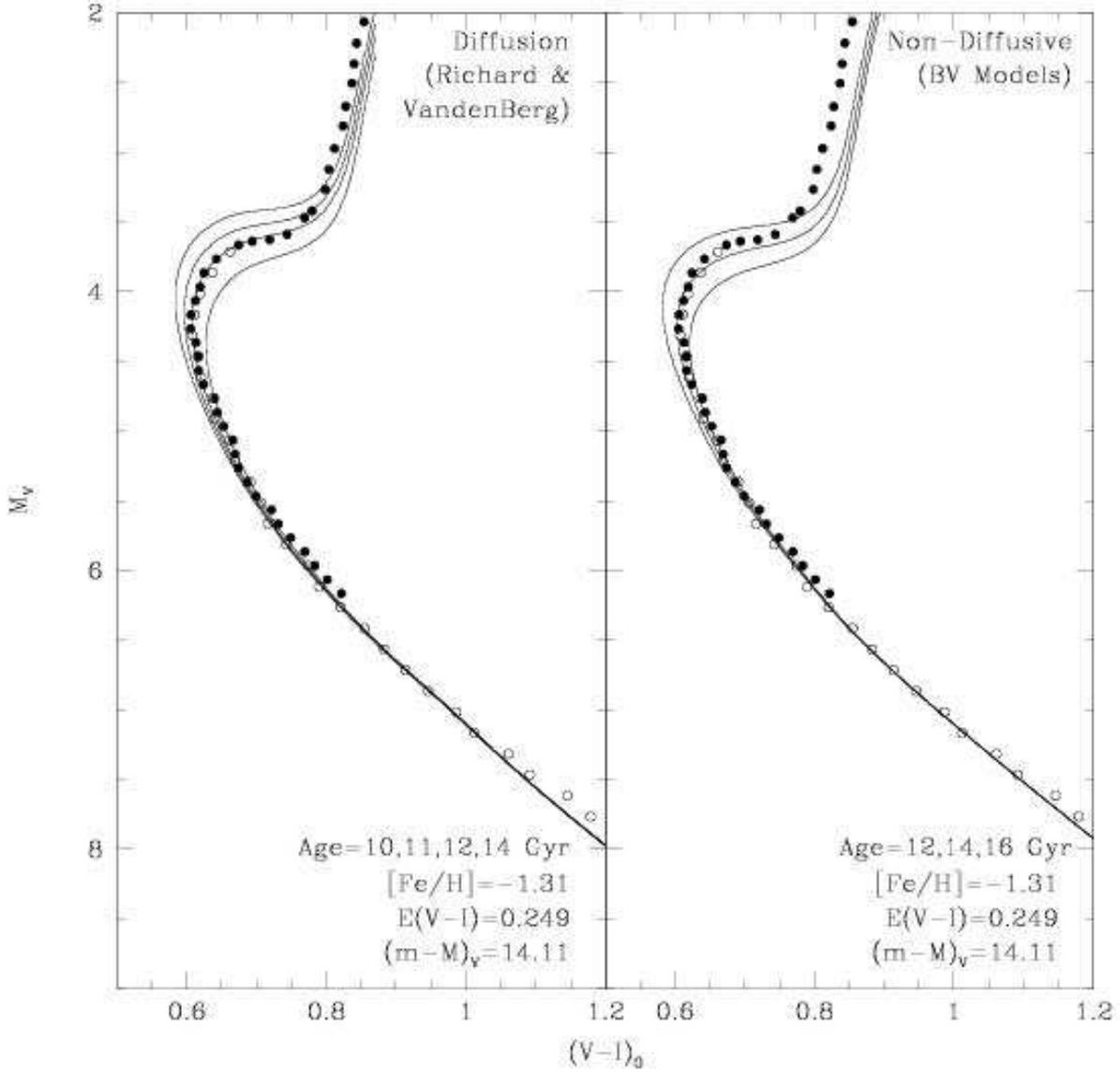}
\caption{Comparison of the fiducial points derived in this study
(\textit{filled circles} are our data; \textit{open circles} are the
\citet{vb02} data) with theoretical isochrones.  No offset has been
applied to the observed colors.
\label{age}}
\end{figure}

\begin{figure}
\plotone{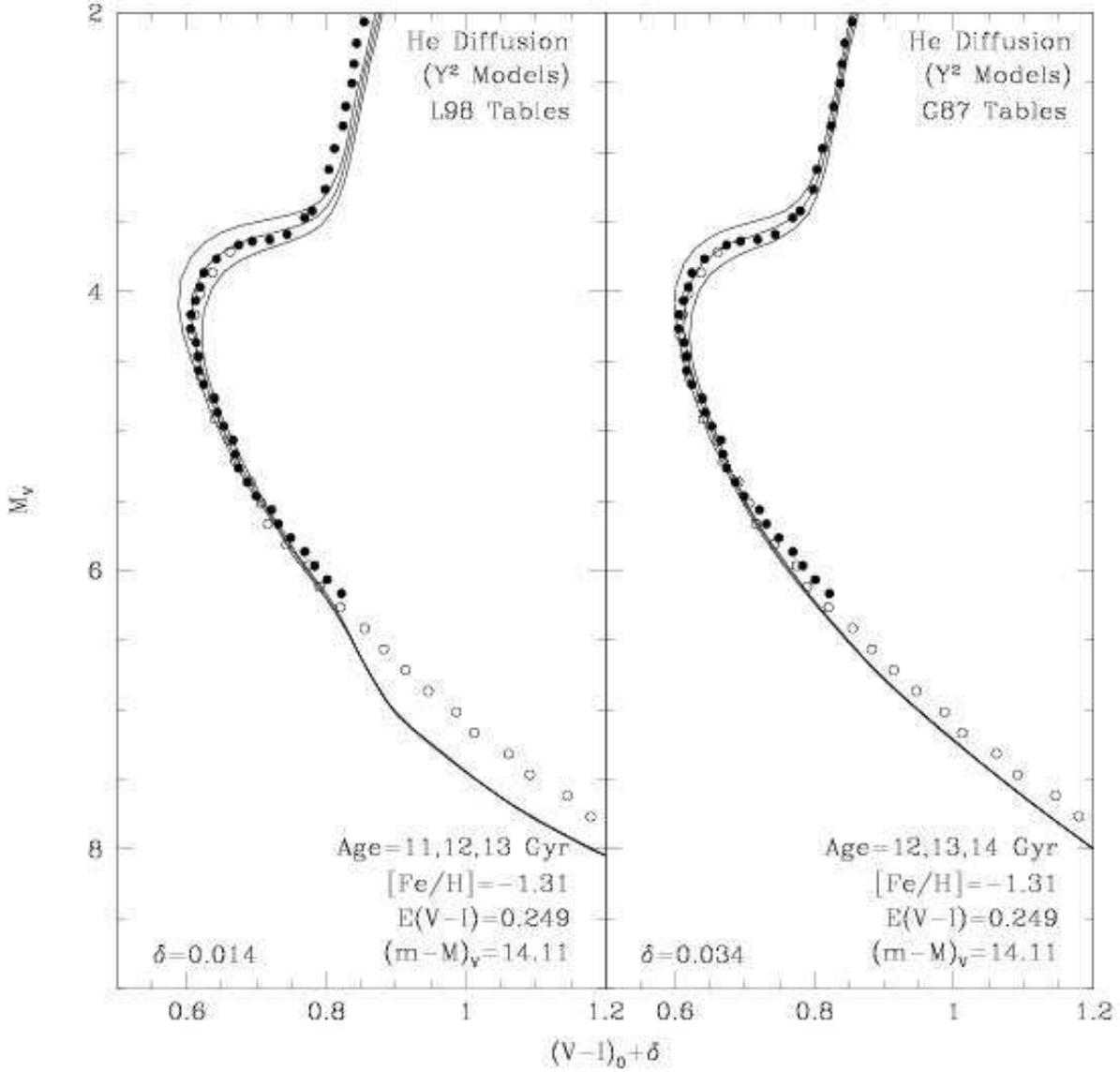}
\caption{Comparison of the fiducial points derived in this study
(\textit{filled circles} are our data; \textit{open circles} are the
\citet{vb02} data) with theoretical isochrones. Offsets of the amounts
$\delta$ have been added to the observed colors to force agreement
on the upper MS.
\label{ageyy}}
\end{figure}

In Figures~\ref{age} and~\ref{ageyy} we plot our fiducial sequences
(using the determined distance modulus and reddening) against the
three models described above.  Following \citet{vb00}, the isochrones have
been shifted by an amount $\delta$ in order to align the colors at the main
sequence turnoff (MSTO). This small shift ($\delta \sim0.02$ for the Y$^2$
models and $\delta=0.001$ for the BV models) accounts for small differences in
color that may arise from photometric zero-point differences, 
$T_{\mbox{eff}}$-color discrepancies, or reddening errors.  Given these 
shifts, we note the inability of the Richard \& VandenBerg, BV, and Y$^2$ 
L98 models to correctly predict the colors of the RGB fiducial sequence.  
For the cluster metal abundance we adopt [Fe/H]$=-1.31$ (close to the CG 
value) in order to make a direct comparison to the Richard \& VandenBerg 
model. Previous work \citep{bv01,va02} has argued for the use of the ZW 
metallicity scale when comparing the BV models to observational data. If we 
assume [Fe/H]$=-1.61$ for the comparison to the BV models, we find that 
one would need an 18 Gyr isochrone in order to match the
fiducial sequence.  Similarly, if we assume this metallicity for the
comparison to the Y$^2$ models, we find that one would need a 16 Gyr L98 Y$^2$
model isochrone or a 16-17 Gyr G87 Y$^2$ model isochrone to match the
fiducial sequence.  These ages are clearly above the recent WMAP upper-limit;
in order to obtain a reasonable age of 13 Gyr given the metallicity on the ZW
scale, the distance moduli would need to be larger by $2-3\sigma$.  Regardless
of the differences in input physics between the BV and Y$^2$ models (and
assuming our values for the distance modulus determined in $\S4.3$),
adoption of the ZW metallicity scale implies an age for M12 that is too old
given recent constraints of the cosmic microwave background measurements
\citep{sp03}. As an estimate of the uncertainty in the deduced ages, we find 
that an error of approximately $\pm0.1$ in $(m-M)_V$ (just below our $1 \sigma$
error) can result in a change of $\pm1.0$ Gyr in age.  Comparisons of the 
observed $V,(B-V)$ fiducial points with these theoretical models implies 
identical ages for M12 to those deduced from Figures~\ref{age} and~\ref{ageyy}.

Given the adoption of the CG metallicity scale, Figures~\ref{age}
and~\ref{ageyy} show that the different models imply slightly different ages
for the cluster. The Richard \& VandenBerg and Y$^2$ models both imply 
reasonable ages ($12-13$ Gyr) for M12, while the BV models imply an older age
by $\sim1-2$ Gyr.  This can most likely be attributed to the inclusion of
gravitational settling in the Richard \& VandenBerg and He diffusion in the
Y$^2$ models and the lack of diffusive physics in the BV models.  In
comparing the Richard \& VandenBerg and BV isochrones, the
diffusive models mimic older non-diffusive isochrones (such as
a shorter SGB) primarily because MS evolution is accelerated by the presence
of additional He in and around the stellar core. This is in agreement with
previous theoretical work done on the effects of He diffusion and GCC ages,
as noted in $\S1$.  Comparing the two Y$^2$ models (see Figure~\ref{ageyy})
it should be noted that because the two Y$^2$ isochrones are calculated from 
identical input physics, the differences between these two can be solely
attributed to the differences in the G87 and L98 $T_{\mbox{eff}}$-color tables. From
the differences between Figures~\ref{age} and~\ref{ageyy} one can
see (not surprisingly) that differences in input physics (Richard \&
VandenBerg vs. BV) and color transformations (Y$^2$ G87 vs. Y$^2$ L98) both
have a significant effect on the comparison of theoretical isochrones with
observed fiducial lines. These differences are of comparable magnitude.  As an
example of the resulting problems, the choice in using the L98 or G87 color 
tables with the Y$^2$ models changes the implied age of M12 (assuming the
correct distance modulus). The left panel of Figure~\ref{ageyy} would imply an
age of $\sim12$ Gyr using the L98 tables, while the right panel (of the same
figure) would imply an age of $\sim13$ Gyr from using the G87 tables.

\subsection{The Luminosity Functions}

\begin{figure}
\plotone{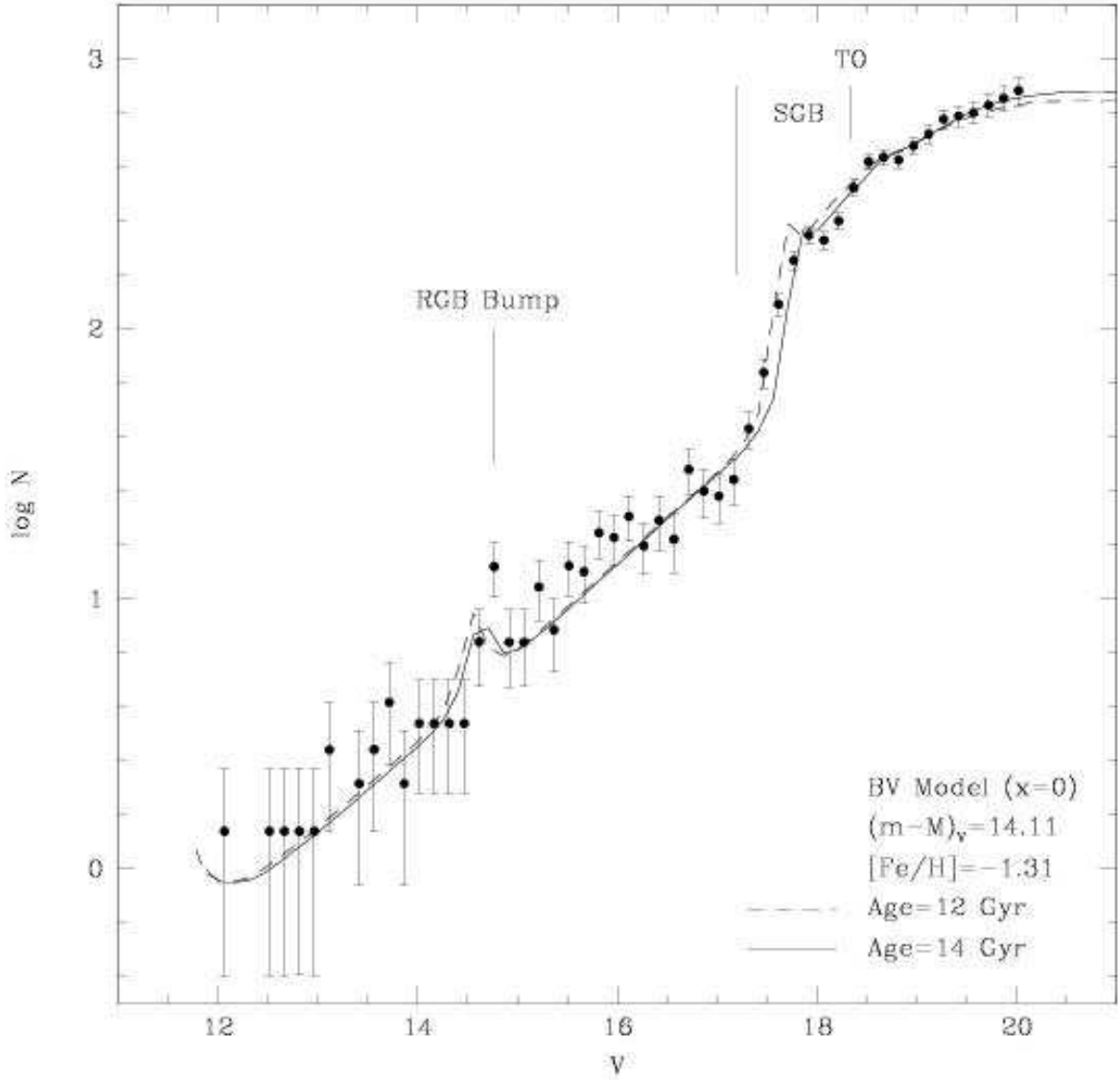}
\caption{Comparison of the observed and theoretical (BV) $V$ band
luminosity function of M12.  The metallicity has been chosen to fall within
the range determined in $\S4.1$, a value close to the CG value. A slightly 
younger model (12 Gyr), or correspondingly smaller distance modulus, shows
better agreement with the SGB ``jump''.
\label{m12vblfv_cg}}
\end{figure}

\begin{figure}
\plotone{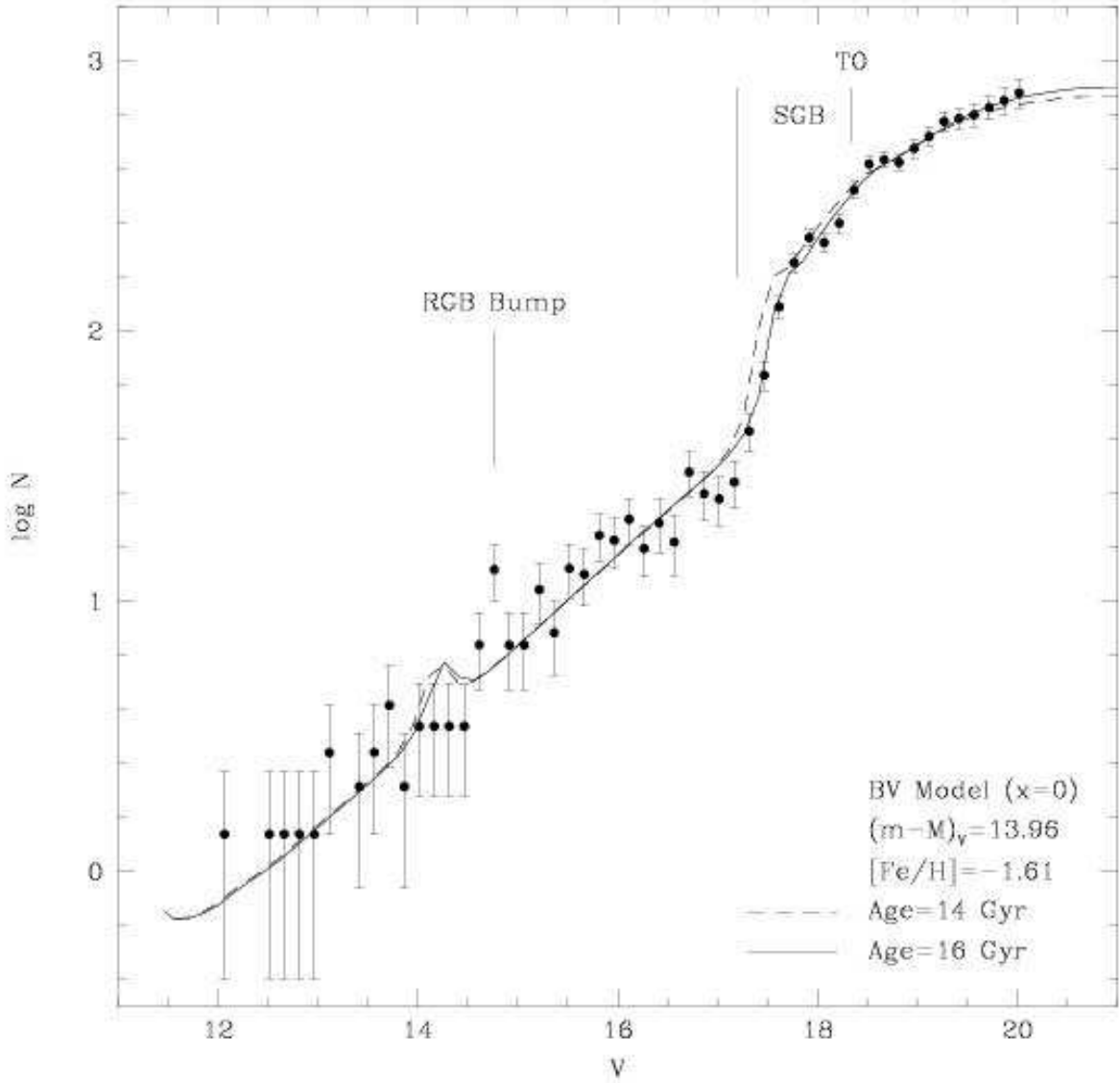}
\caption{Comparison of the observed and theoretical (BV) $V$ band
luminosity functions of M12.  The ZW metallicity value has been chosen
following the discussion in $\S6.1$ and $\S6.2$.  A slightly older
model (16 Gyr), or correspondingly larger distance modulus, provides a
better description of the observed data.
\label{m12vblfv}}
\end{figure}

\begin{figure}
\plotone{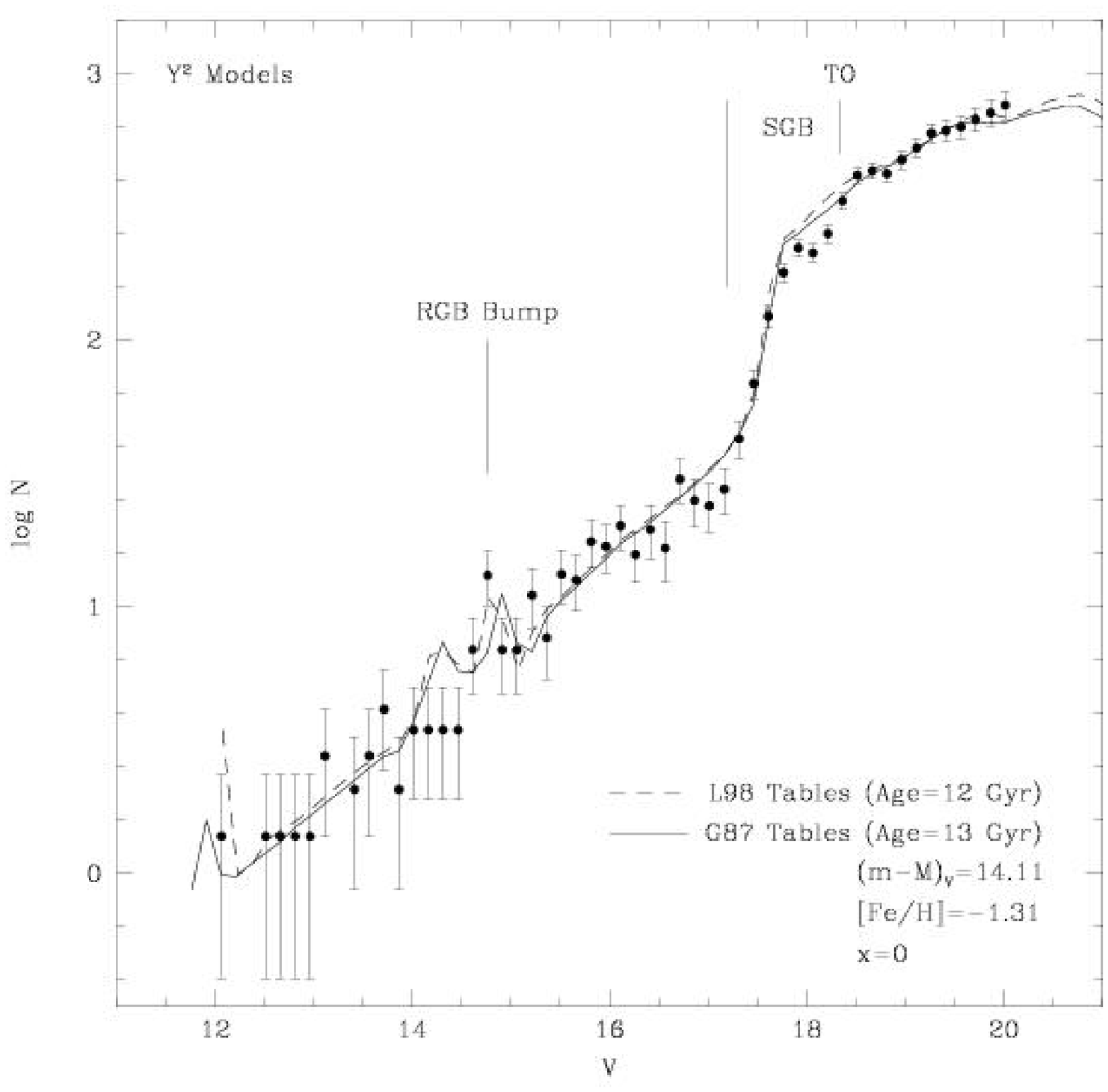}
\caption{Comparison of the observed and theoretical (Y$^2$) $V$ band
luminosity functions of M12 for the two different ages implied by the
G87 and L98 color transformation tables in Figure~\ref{ageyy}.}
\label{m12yylfv}
\end{figure}

\begin{figure}
\plotone{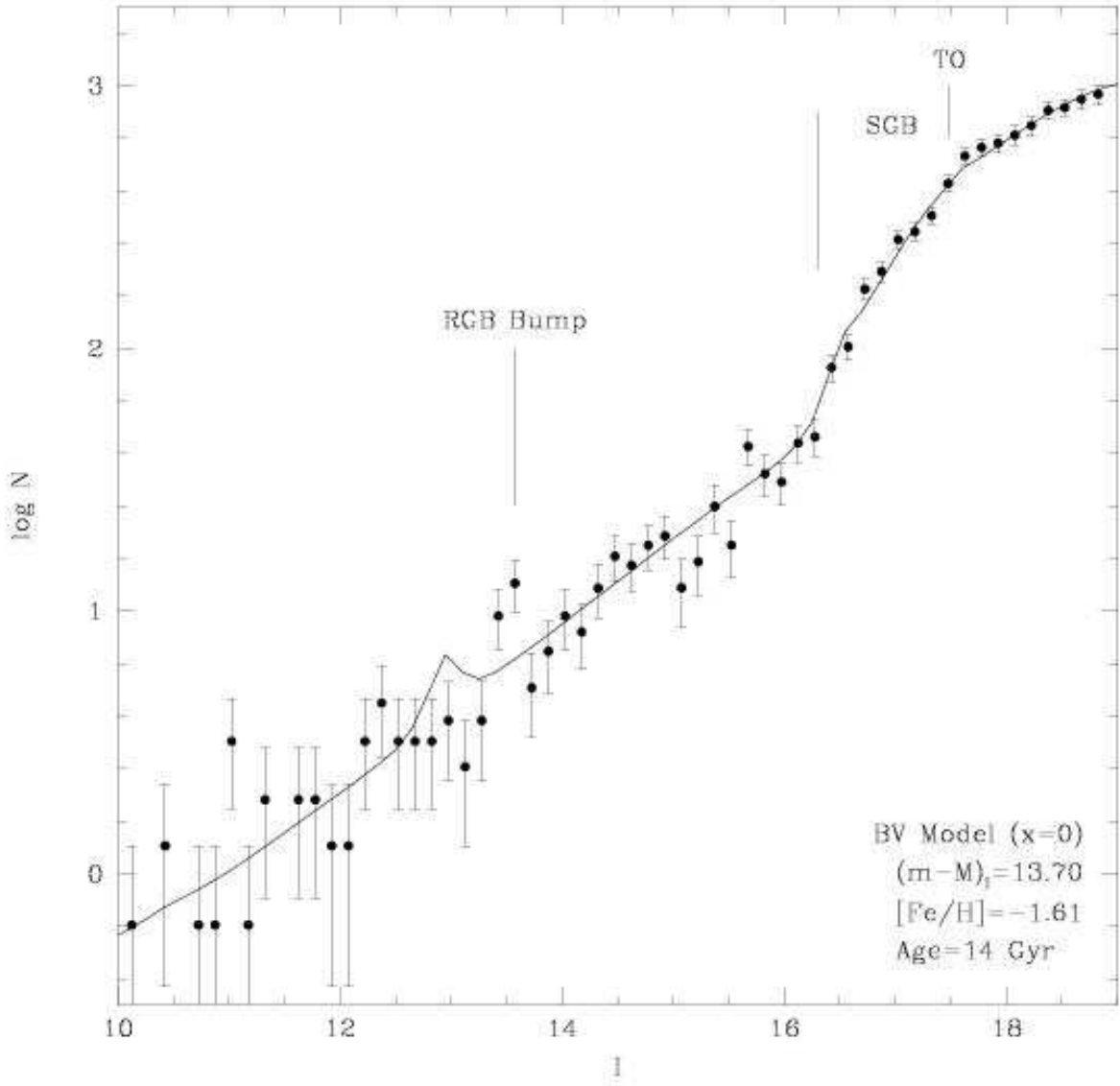}
\caption{Comparison of the observed and theoretical (BV) $I$ band
luminosity functions of M12.
\label{m12vblfi}}
\end{figure}

\begin{figure}
\plotone{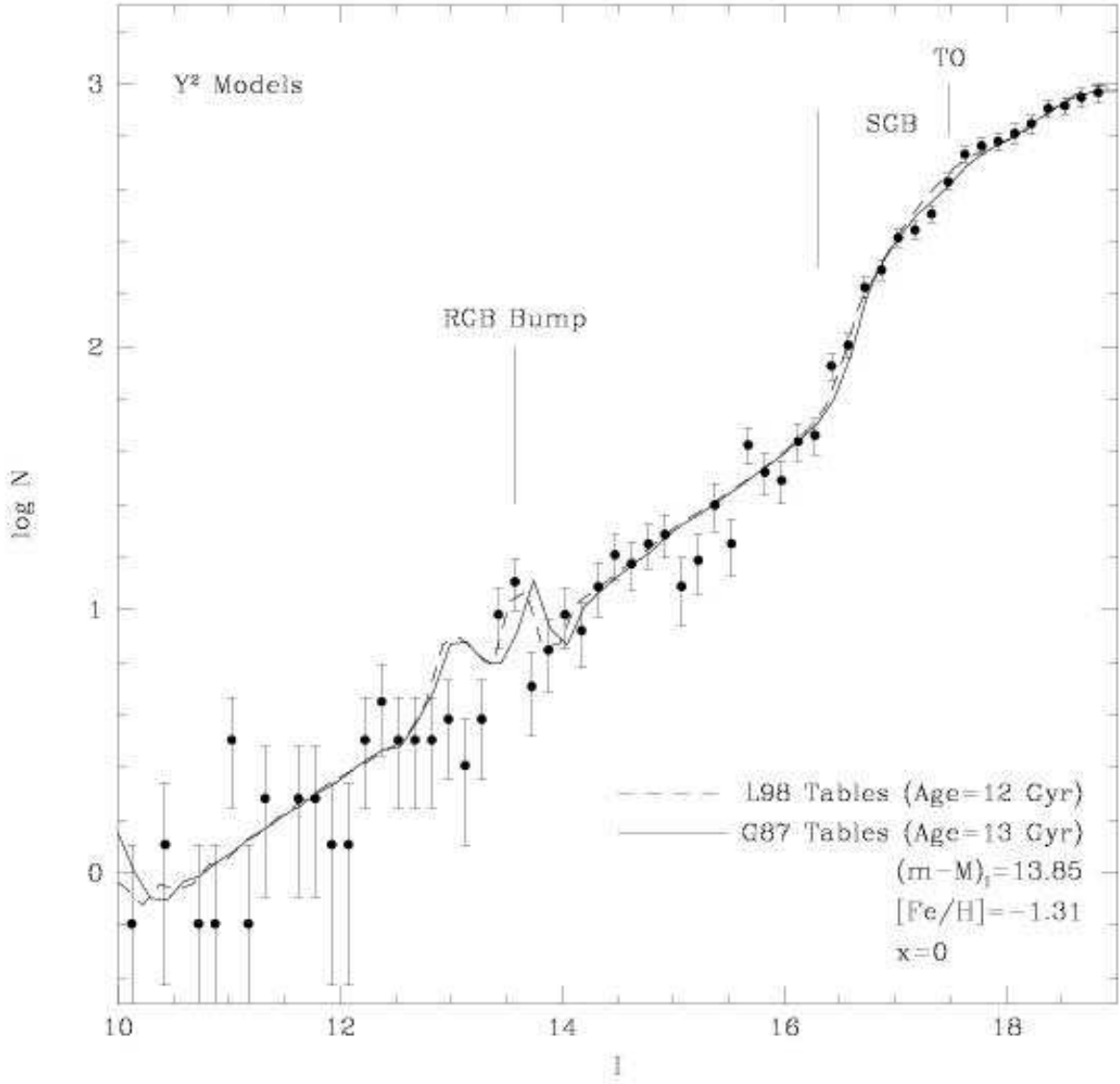}
\caption{Comparison of the observed and theoretical (Y$^2$) $I$ band
luminosity functions of M12 for the two different ages implied by the
G87 and L98 color transformation tables in Figure~\ref{ageyy}.
\label{m12yylfi}}
\end{figure}

We compare the observed $V$ and $I$ LFs for M12 to theoretical
LFs in Figures~\ref{m12vblfv_cg}-\ref{m12yylfi}.  In
doing this, we wish to investigate whether the physics used in the
theoretical models can adequately explain our observations.
Theoretical LFs were generated for the BV and Y$^2$ models only,
since LF data were not available from the Richard \& VandenBerg models
(D. A. VandenBerg 2003, private communication).  Given the previous
discussion over the metallicity scale and BV models (see $\S6.1$) we
show two comparisons of our observed $V$ band LF with the BV models, one
for a cluster metallicity near the CG scale (Figure~\ref{m12vblfv_cg}) and
one for a metallicity on the ZW scale (Figure~\ref{m12vblfv}).  For both
these comparisons we show a 14 Gyr model as implied by the CMD in
Figure~\ref{age}. The LFs for this age should mimic younger diffusive models 
(see Proffitt \& VandenBerg 1991 Figure 18).  Despite the choice of metal
abundance in the BV models we find that an age adjustment (or
correspondingly a distance
modulus change) is necessary in order to match the SGB ``jump'' in the $V$
band LF, although the adjustment is younger in one case (the CG 
metallicity; Figure~\ref{m12vblfv_cg}) and older in the other (the ZW 
metallicity; Figure~\ref{m12vblfv}).  The older age necessary for the ZW
comparison is in agreement with the discussion in the previous
section regarding the ZW metallicity scale; an older age (or larger distance 
modulus)  to provide a better description of the data.  

For LF comparisons with the Y$^2$ models (Figures~\ref{m12yylfv}
and~\ref{m12yylfi}), we adopt the metallicity of 
[Fe/H]$=-1.31$ because of the agreement of the
Y$^2$ isochrones with the observed CMD for evolved stars (see
Figure~\ref{ageyy}).  In the $I$ band the shape of the Y$^2$ LFs is 
somewhat dependent on the choice of $T_{\mbox{eff}}$-color transformations, 
particularly in those regions where the CMD is changing the most (e.g. 
the MSTO and SGB regions).  The $V$ band LF does not depend on the 
$T_{\mbox{eff}}$-color relations but only on the bolometric corrections.  
The various $V$ band distance moduli in 
Figures~\ref{m12vblfv_cg}-\ref{m12yylfi} were determined via
subdwarf fitting (as in $\S4.3$) assuming the value specific to the
metallicity used for the theoretical model. The theoretical LFs were
normalized (over a range of 0.3 mag) to the total number of stars in the
M12 LF sample at a point on the upper MS $\sim1$ magnitude fainter than
the MSTO.  A mass function exponent $x=0$ (where $N(M)\propto
M^{-(1+x)}$) was selected as to match both the $V$ and $I$ band LFs at
the faint end of the sample. This mass function exponent is in agreement
with the one found by S89.

\subsubsection{The Subgiant Branch}

As noted in the introduction, some metal-poor clusters have shown evidence
for an excess of stars on the SGB portion of the LF.  In general the
observed $I$ band SGB LF of M12 shows better agreement with theory than
does the $V$ band SGB LF, which is noticeably ``jagged'' compared to the 
``smooth'' theoretical models.  Given that we have eliminated the faintest
stars ($V<16$) in the core region of the cluster ($r<200$ pixels = 2\Min3),
it is unlikely that stellar blends can account for the discrepant SGB LF.
To test this possibility, we computed the $VI$ LFs for a restricted region
of the data.  After eliminating the core region ($r<250$ pixels = 2\Min8) we 
found no substantial difference in the observed LFs, thus justifying our
radial cut in the LF computation.  In our examination of the SGB region of
the M12 LF we apply three different techniques in order to investigate the
discrepancies between observations and theory:

\begin{itemize}
\item First we compare the theoretical and observed SGB LFs in an
``absolute'' fashion, using the values derived in $\S4$ for the cluster
parameters.

\item Second, we make the SGB LF comparison after performing a shift to
bring a common point on the upper MS into coincidence. 

\item Third, we formulate a technique to maximize the exploration of the
evolutionary timescales of the cluster stars by selecting theoretical
models based on compatibility with the observed CMD.  
\end{itemize}

Lastly, we construct the LF of M12 from the \textit{HST} data of \citet{pi02} 
and compare this result with the LF derived in this study.

\begin{figure}
\plotone{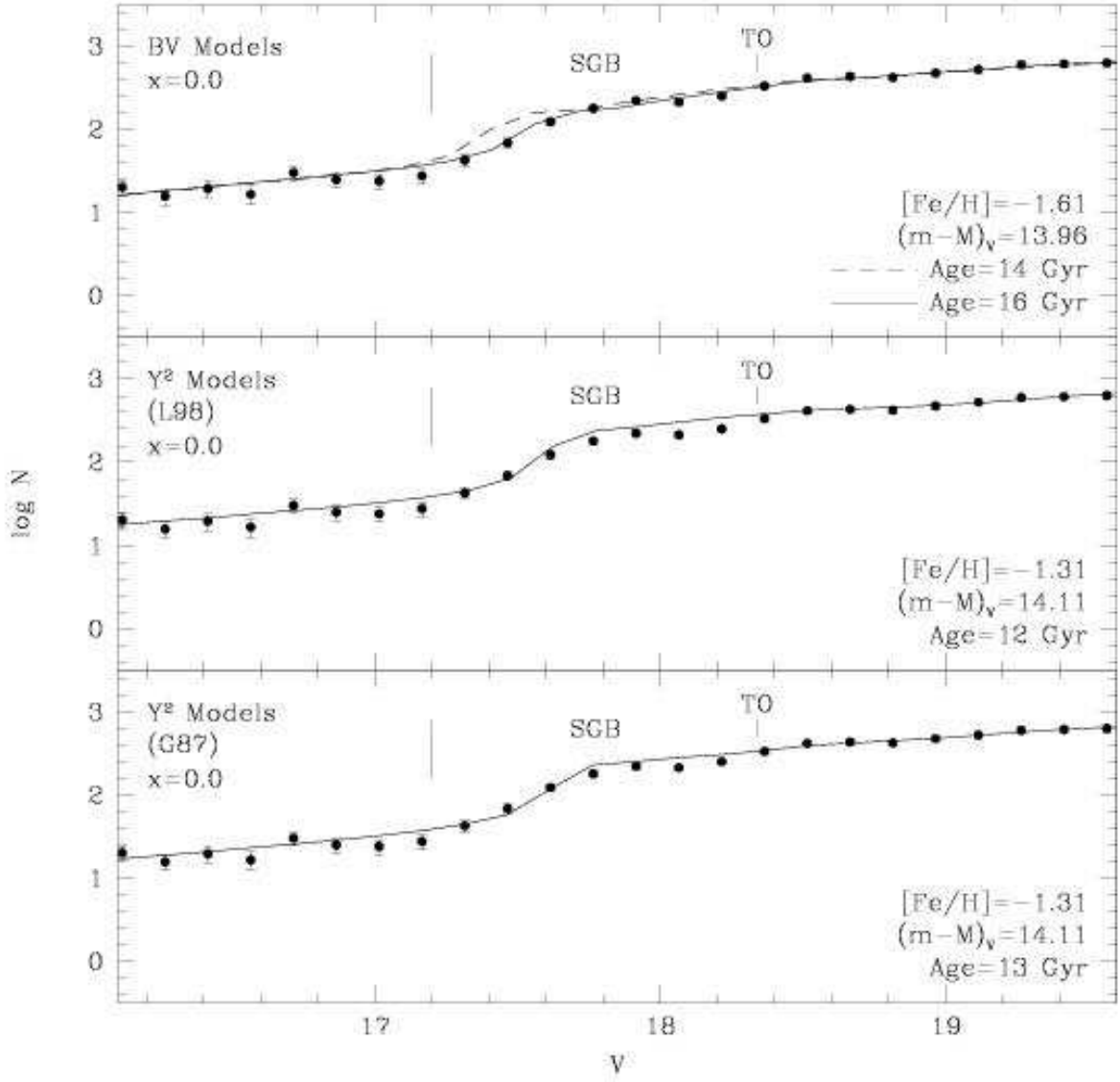}
\caption{SGB region of Figures~\ref{m12vblfv} and~\ref{m12yylfv}, 
showing the M12 $V$ band luminosity function with the BV and Y$^2$ 
theoretical models.
\label{sgbcompvabs}}
\end{figure}

\begin{figure}
\plotone{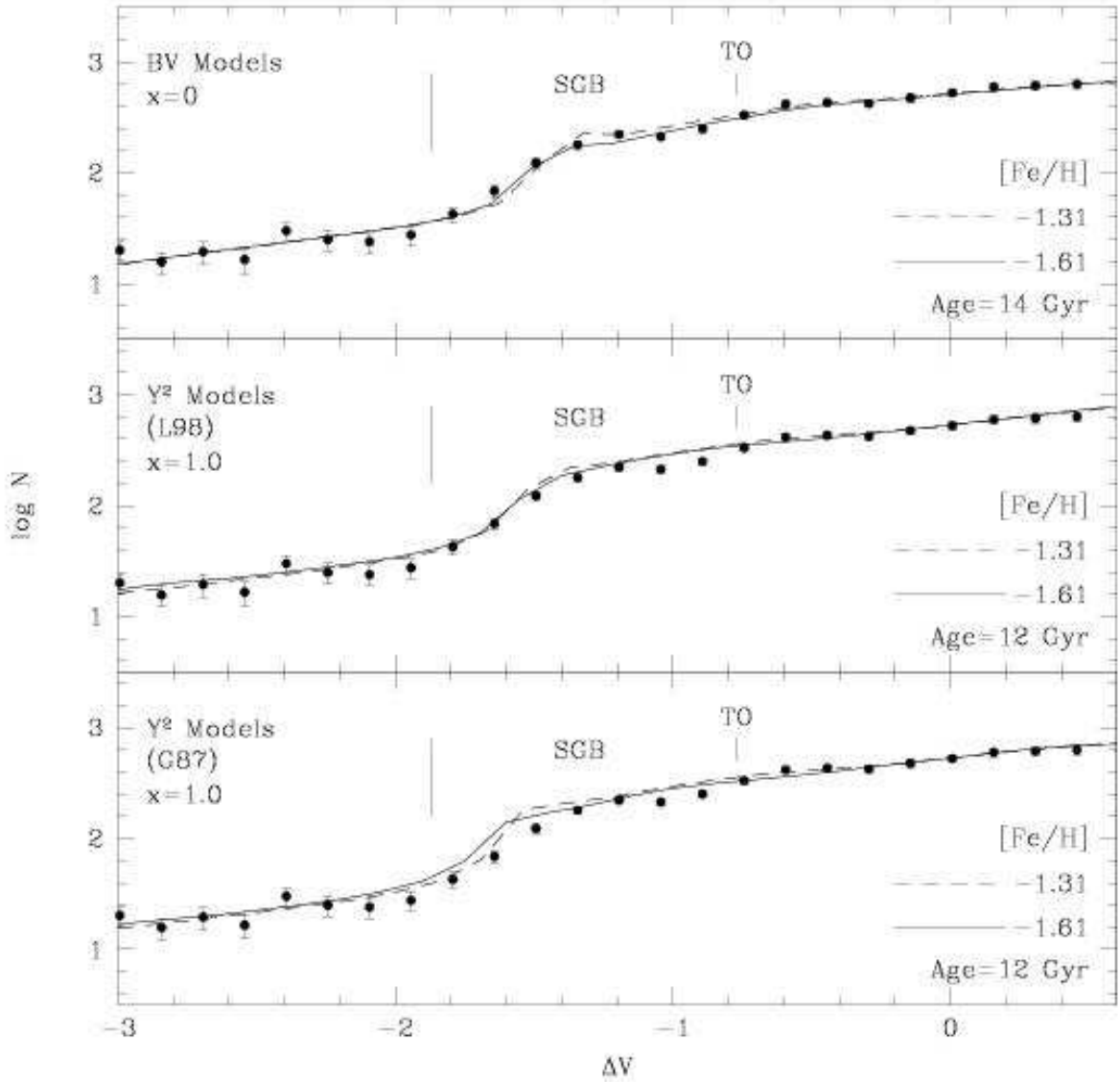}
\caption{SGB region of the M12 $V$ band luminosity function shown
with the BV and Y$^2$ theoretical models for two values of the
cluster metallicity. The magnitude scale has been shifted to match a
common point on the upper main-sequence as discussed in $\S6.2.1$.  The middle
and bottom panels are identical except for the choice of color-$T_{\mbox{eff}}$
transformation table.
 \label{sgbcompv2}}
\end{figure}

In Figure~\ref{sgbcompvabs} we highlight the SGB and upper MS region of the
M12 $V$ band LF with the same theoretical models from Figures~\ref{m12vblfv} 
and~\ref{m12yylfv}. These comparisons use the parameters derived in $\S4$,
where the ages were adopted from the CMD analysis in $\S6.1$.  While the BV
models appear to give the best overall description of the observed SGB
region, this is only true if an older age (or larger distance modulus) is
adopted (as noted in $\S6.1$ and $\S6.2$).  In these ``absolute''
comparisons, both the Y$^2$ models predict more stars than are observed in
the SGB LF bins between the SGB ``jump'' and the MSTO. Figure~\ref{sgbcompv2} 
shows the observed SGB region of the M12 $V$ band LF again, but here we
use the second technique (listed above) for the comparison to theory.
Figure~\ref{sgbcompv2} also shows theoretical LFs for metallicities on both 
the CG and ZW scale. In the theoretical models the SGB LF shapes are
largely due to the choice of metallicity and age; the strongest dependence
is usually on metallicity
\citep{zo00}.  However, the choice of $T_{\mbox{eff}}$-color
transformations also has a noticeable impact on the shape of the theoretical
SGB LF. The Y$^2$ model comparisons in Figure~\ref{sgbcompv2}
are identical except for the choice of color transformation table.  Because
of the strong correlation between metallicity, distance modulus, and age we
shift the magnitude scale (of both the theoretical and observed CMDs) to
bring a point on the upper MS into coincidence.  We follow the method
described by \citet{st91}, using as reference the point on the upper MS that
is 0.05 magnitudes redder than the MSTO magnitude.
In this formalism, the distance modulus has been eliminated and hence
the age and metallicity will be difficult to determine (in an absolute
sense) in such diagrams \citep{st91}. Uncertainties in the
zero-pointing could be as high as $\sim0.1$ mag, mostly because the
slope of the upper MS differs between sets of isochrones. From
Figure~\ref{sgbcompv2} we see that no choice of metallicity, color table,
and model is able to completely describe the observed SGB LF.  The BV and
Y$^2$ models are able to match some points for a higher metallicity leaving
other points lower than predicted, but a lower metallicity will mean other
points are \textit{higher} than predicted. Using the Y$^2$ models, a better
description of the SGB region (that closely resembles the BV comparisons)
can be found if a slightly larger mass function exponent $x=1$ is adopted
(compare the SGB regions in Figures~\ref{sgbcompvabs} and~\ref{sgbcompv2}).

\begin{figure}
\plotone{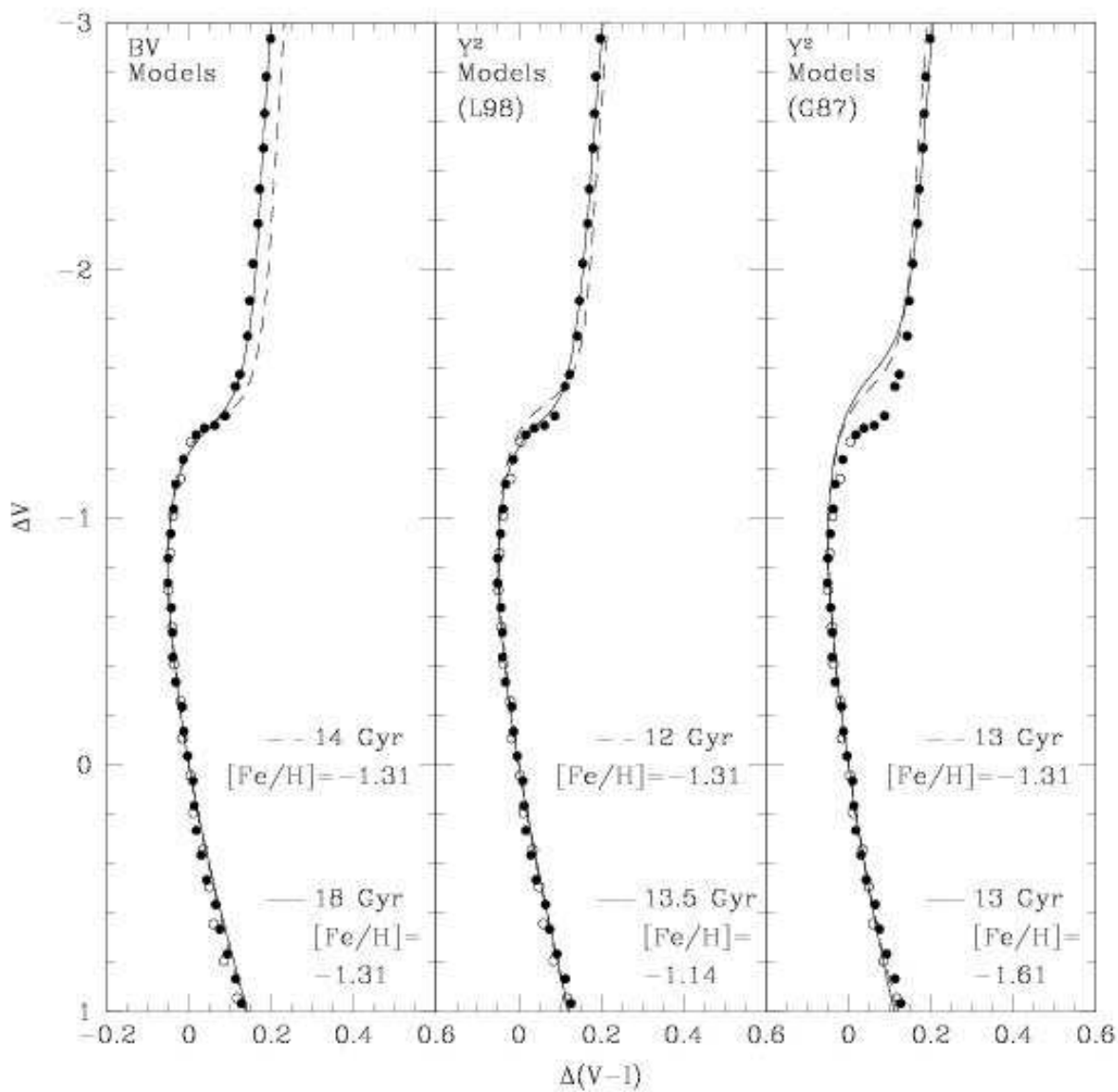}   
\caption{Comparison of the observed fiducial sequence of M12 with the BV
(\textit{left panel}) and Y$^2$ (\textit{middle} and \textit{right panels})
theoretical isochrones.  Both the magnitude and color scales have been
shifted to match a common point on the upper main-sequence as discussed in
$\S6.2.1$.  The ``best fit'' isochrones from Figures~\ref{age} and~\ref{ageyy}
are shown as the \textit{dashed line}, but we find the \textit{solid line}   
theoretical models to better match the observations when the CMD is shifted  
in this manner.  We are unable to find a Y$^2$ theoretical model
(using the G87 color table and within the deduced range of cluster
metallicity) that adequately describes the observed CMD.
\label{age_all}}
\end{figure}

\begin{figure}
\plotone{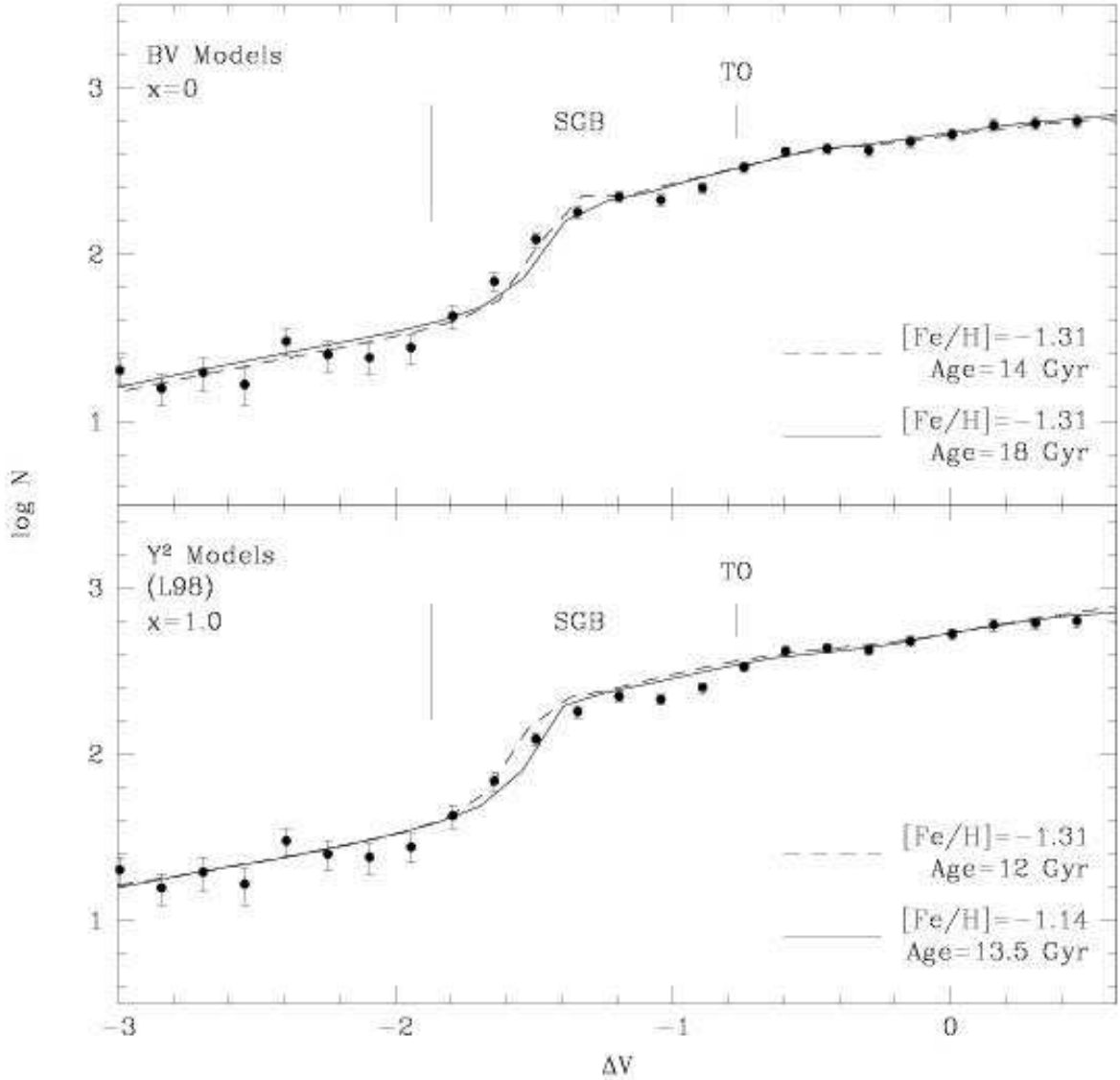}
\caption{Comparison of the observed $V$ band LF of M12 and the BV and Y$^2$
(using the L98 color-$T_{\mbox{eff}}$ transformation table) theoretical LFs
corresponding to the ``best fit'' isochrones in Figure~\ref{age_all}
(\textit{solid line}). The magnitude scale has been shifted to match a
common point on the upper main-sequence as discussed in $\S6.2.1$. 
The (\textit{dashed lines}) theoretical models are identical to the
\textit{dashed line} theoretical models from Figure~\ref{sgbcompv2} and are
plotted for comparison purposes.
\label{sgbcompviso}}
\end{figure}

Systematic errors in the $T_{\mbox{eff}}$-color transformations can
lead to distortions of the theoretical isochrone in the observational
CMD, and can thereby affect the theoretical LFs where the isochrone is
changing most quickly in color. Independent of that, the slope of the
subgiant branch in the theoretical HR diagram is affected by cluster
parameters like age and metallicity. In order to minimize systematic
differences between theoretical and observed LFs, and to attempt to
focus on the evolutionary timescales of the stars, we devised another
method of making the comparisons. In this method, we selected the
theoretical isochrone that best matched the observed CMD fiducial
sequence \textit{when shifting the magnitude and color scales to match
a point on the upper MS} (as described above). The theoretical model
that best describes the observed CMD will be different depending on
whether or not one performs this shift (or uses the determined
distance modulus and reddening).  While this technique places significant
weight on the ability of the $T_{\mbox{eff}}$-color tables to correctly
describe observations, it is unlikely that physical models will
provide a good description of the observed LF if the theoretical and
observed CMDs do not match.  In Figure~\ref{age_all} we show the observed
M12 fiducial sequence with the BV and Y$^2$ models when shifting the color
and magnitude scale.  The theoretical model that best described the
observations using the ``absolute parameters'' in \S 6.1 (the {\it dashed
line} in Figure~\ref{age_all}) is shown with another model that better
describes the fiducial sequence when
performing this shift (the {\it solid line} in Figure~\ref{age_all}).
We are unable to find an adequate theoretical description of the CMD
observations using the G87 color table with the Y$^2$ model (given the
range of metallicity determined in \S 4.1).  While this throws some
doubt on the ability of the G87 tables to match observed colors, this
should be further investigated for other GGC CMD observations.  The
metallicities required for the BV and Y$^2$ models are within the
range of M12 observations previously quoted (\S 4.1), but the age of
the BV model must be very large to match the shape of the SGB region
of the fiducial line.  Using the ``best fit'' BV and Y$^2$ L98
descriptions (from Figure~\ref{age_all}), we show the observed M12 SGB
LF with the theoretical LFs generated from these models in
Figure~\ref{sgbcompviso}.  The theoretical LFs using our determination
of the ``absolute'' parameters (i.e.  the same theoretical LFs from
Figure~\ref{sgbcompv2}) are also shown.  Neither set of models
provides entirely adequate descriptions of the SGB region of the M12
LF. For the BV models, the SGB evolutionary timescale is somewhat
underestimated. For the Y$^2$ models, the numbers of stars in the two
magnitude bins just brighter than the MSTO point are more noticeably 
over predicted, and the slope of the LF for the bright SGB is predicted too
steep.

\begin{figure}
\plotone{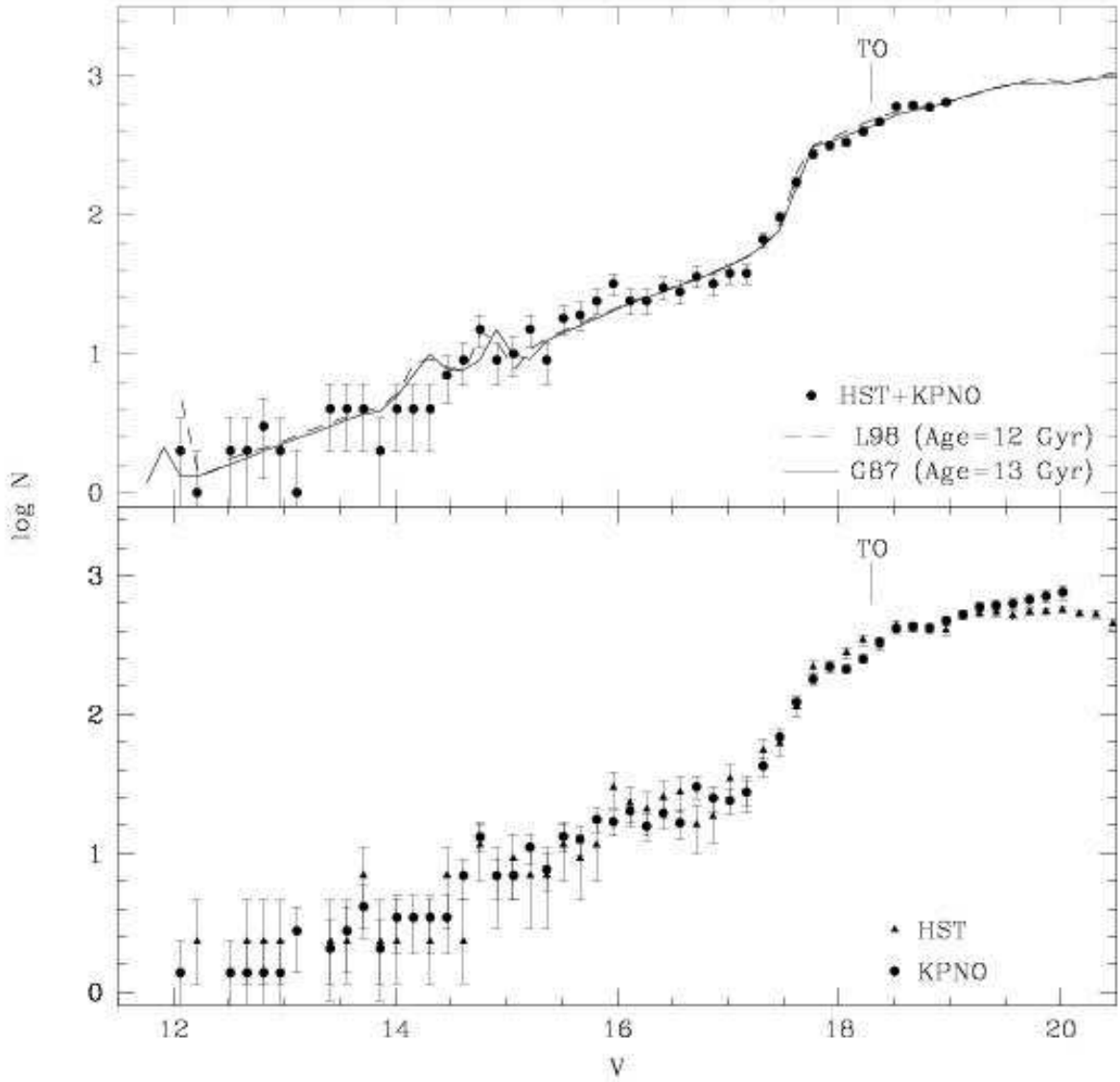}
\caption{\textit{Top Panel}: Luminosity function of M12 formed from
combination of the \textit{HST} data \citep{pi02} and the KPNO data
(this study).  \textit{Bottom Panel}:  The \textit{HST} and KPNO data
separately, with the \textit{HST} data scaled to match the KPNO data at
the upper main-sequence.
\label{hstkpno1}} 
\end{figure}

To further investigate the presence of over- or under-abundances of
stars in the SGB region, we construct an M12 LF using a combination of
our wide-field data with the \textit{HST} photometry of \citet{pi02}.  This
summation of data sets will increase the statistical significance of the
LF, while examination of the data set separated will allow for the
inspection of the LF at differing radii from the cluster center.  
Figure~\ref{hstkpno1} shows both the LFs separately and the combined LF
compared with the Y$^2$ theoretical models of Figure~\ref{m12yylfv}. We
only compute the combined LF down to $V=\sim19$ in order to avoid
complications with incompleteness in the \textit{HST} data set.  Two 
magnitude bins stand out when comparing the \textit{HST} and KPNO data sets
separately; the two bins just brighter than the MSTO appear to have more
stars in the \textit{HST} data than the KPNO data.  If it is a ``real'' 
effect (e.g. not a computational artifact), then this would imply a 
larger number of SGB stars concentrated towards the cluster center.  As can
be seen in the combined (\textit{HST}+KPNO) LF, the SGB region still shows 
a slight underabundance of stars compared to theoretical predictions. 
However, near $V=18.5$ (just fainter than the MSTO) there appears to be a
small increase in the number of stars.  In general, these discrepancies are
small and at best statistically significant at $\sim1\sigma$ level.  As a
judge of the goodness-of-fit, we compute the reduced $\chi^2$ (denoted
$\chi_{\nu}^2$) for the two Y$^2$ models in Figure~\ref{hstkpno1}.  For the
12 Gyr L98 Y$^2$ model we find $\chi_{\nu}^2=1.77$ and for the 13 Gyr G87
Y$^2$ model we find $\chi_{\nu}^2=1.55$.  As a comparison between the KPNO
and \textit{HST} data, we find $\chi_{\nu}^2=2.85$.  If we only compute
$\chi_{\nu}^2$ for the MSTO and SGB regions of the LF ($17.3<V<19.0$ we
find $\chi_{\nu}^2=4.57$ as a comparison between the KPNO and \textit{HST}
data, $\chi_{\nu}^2=2.99$ for the 13 Gyr G87 Y$^2$ model, and 
$\chi_{\nu}^2=4.07$ for the 12 Gyr L98 Y$^2$ model.  In summary, the LF
formed from the inclusion of the \textit{HST} data set with our wide-field
data still shows a discrepant SGB LF compared to theory.

\subsubsection{The Red Giant Branch}

The theoretical and observed $VI$ band RGB LFs appear to be in
agreement within the errors for both the BV and Y$^2$.  We note the
detection of the RGB ``bump'' (in both the cumulative and differential
LFs) at $V\sim14.7$, $I\sim13.6$.  This is consistent with the $V$
magnitude of the bump determined by \citet{fe99}. The observed $I$
band RGB LF agrees well for the most densely-populated portion fainter
than the RGB bump ($13.4>I>16.2$, the ``lower'' RGB). Comparisons
between the theoretical models show that the RGB slopes appear to be
in good agreement. Noting the predicted and observed numbers of
MS-to-RGB stars, there appears to be no discrepancy within the errors;
M12 does not appear to have an excess of RGB stars compared to theory.

Both the BV and Y$^2$ models predict the presence of the RGB bump.  The
standard interpretation of the RGB bump is that it is due to the
movement of the hydrogen burning shell through the chemical composition
discontinuity left by the deepest penetration of the envelope convective 
zone \citep{th67,ib68}.  The Y$^2$ models show two peaks near the
observed RGB bump, but the brighter of the two peaks is the true RGB
bump. The fainter, larger peak is a numerical artifact of the models
due to luminosity grids which are sparse and non-uniform in this region
of the models (S. Yi, private communication).  Thus, both the BV and
Y$^2$ predict a RGB bump at very similar magnitudes ($V=\sim14.2$) but
are $\sim0.2$ magnitude brighter than actually observed.  While this 
absolute comparison of the bump position disagrees with theory, recent 
work by \citet{ri03} has shown the bump position \textit{relative to the 
horizontal branch} (for their sample of 54 GGCs) is in agreement with 
their most recent stellar evolution models.

\subsection{Comparison of Theoretical and Observed LFs for Other Clusters}

\begin{figure}
\plotone{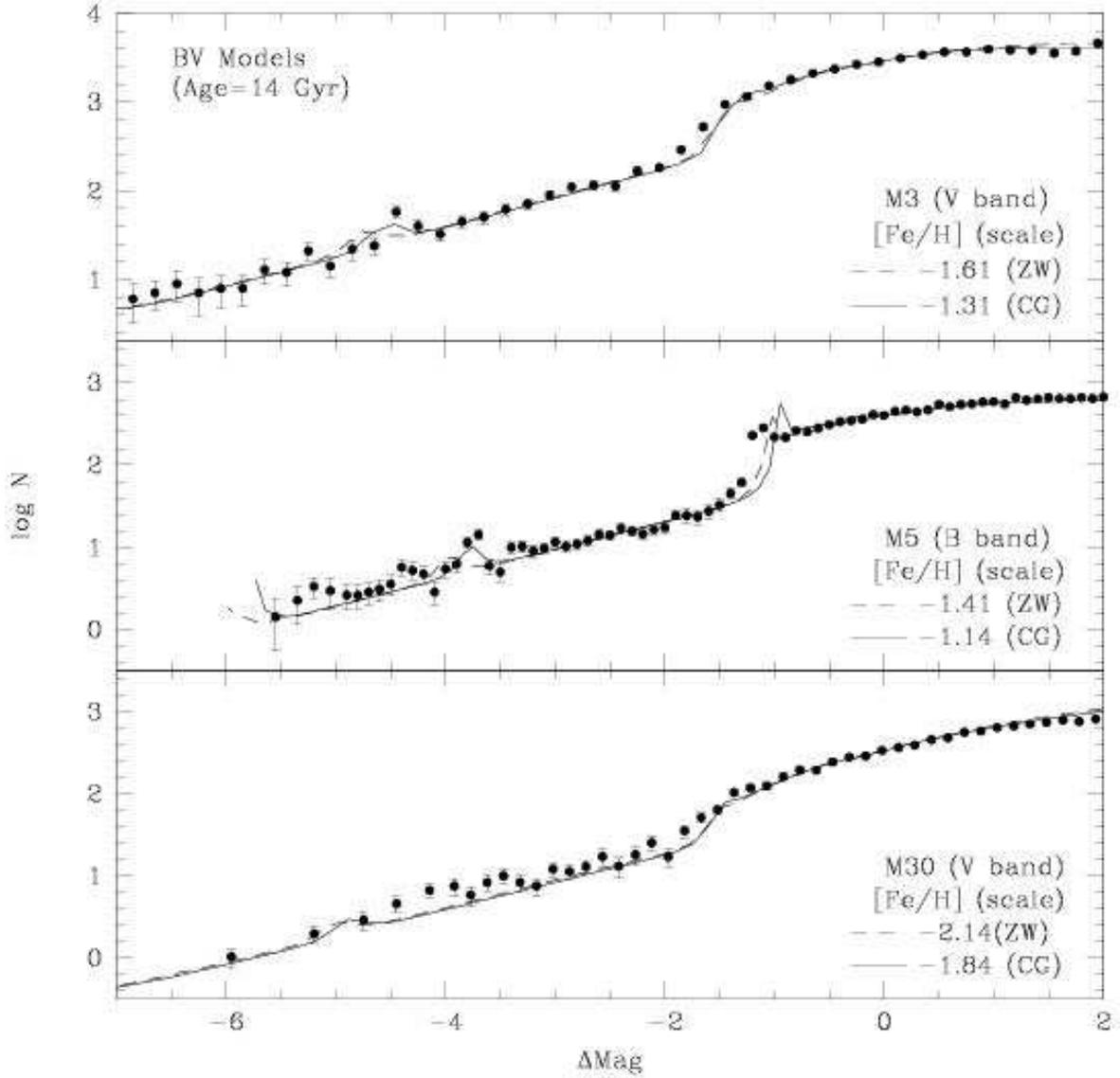}
\caption{Comparison of the observed luminosity functions of M3 (\textit{top 
panel}), M5 (\textit{middle panel}), and M30 (\textit{bottom panel}) with the
BV theoretical luminosity functions.  The magnitude scale has been shifted to
match a common point on the upper main-sequence as discussed in $\S6.2.1$.
\label{vblfz}}
\end{figure}

\begin{figure}
\plotone{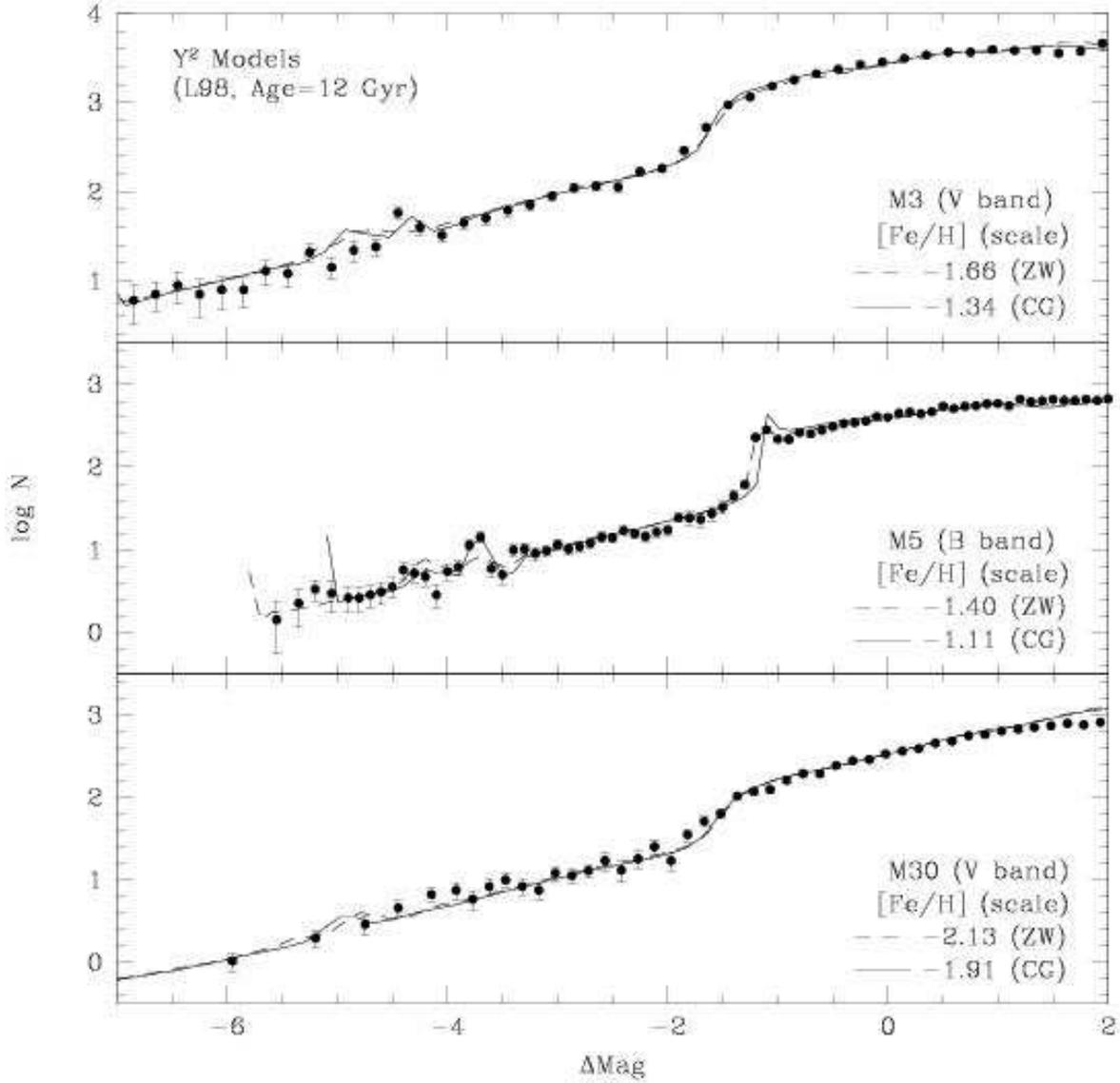}
\caption{Comparison of the observed luminosity functions of M3 (\textit{top 
panel}), M5 (\textit{middle panel}), and M30 (\textit{bottom panel}) with the  
Y$^2$  theoretical luminosity functions (using the L98 color-$T_{\mbox{eff}}$
transformation table). The magnitude scale has been shifted to
match a common point on the upper main-sequence as discussed in $\S6.2.1$.
\label{yylfz}}
\end{figure}

Because of the influence of cluster metal abundance on the SGB
morphology, we compare the BV and Y$^2$ theoretical models to the LFs
of three other well-studied GGCs. In doing this we seek to make an
attempt to determine whether one set of models can reproduce the main
features of the LFs of clusters covering a wide range of
metallicities. Figures~\ref{vblfz} and~\ref{yylfz} compare the
observed LFs of M3 \citep{rood99}, M5 \citep{sand96}, and M30
\citep{sa99} to theoretical LFs with metallicities from the ZW and CG
scales.  The mass function exponents were taken from the referenced
studies: $x=0$ for M3, $x=0.5$ for M5, and $x=2.0$ for M30.  The
magnitude scale was shifted to a common zero-point as described above
to eliminate the sensitivity to distance modulus and age.  While this 
comparison does present clusters having a wide range of central densities,
there is no evidence for population gradients in normal clusters and
some evidence for this effect in post-core-collapse clusters (such as M30;
see \citet{bu96}).  Given the radial cuts necessary to remove poorly-measured
stars from the central regions of the clusters, the presence of crowding and 
possible population gradients should have a negligble influence on the
shape of the cluster LFs.

The plots indicate that the SGB region is also somewhat insensitive to
metallicity except for filter choices which the cause the SGB to be
nearly horizontal, as is the case for the $B$-band LF of M5 (Sandquist
et al. 1996). In the case of M5, the Y$^2$ theoretical LF using the ZW
scale value is the most consistent with the observations. This feature
might be exploited in future LF studies to help nail down the absolute
metallicity scale. However, because the SGB is a feature primarily
involving $T_{\mbox{eff}}$ change, current uncertainties in the
$T_{\mbox{eff}}$-color transformations would have to be removed first.
A comparison between the G87 and L98 tables for these clusters shows
better agreement with observations when using the L98 transformations,
as is also confirmed in comparing the middle and bottom panels of
Figure~\ref{sgbcompv2}.  For all 3 clusters (and for M12 also), the BV models 
are unable to match the SGB jump and hence we are unable to choose between
metallicity scales using these models.

Though the ZW scale is favored in these {\it relative} comparisons
between theory and observations, in an \textit{absolute} sense, the
choice of the ZW metallicity scale in the M12 analysis means that the
determined distance modulus will be too small to match our LF
observations. (This is consistent with the discussion of the CMD
$\S6.1$.) The ZW scale implies that the use of an older model (or
larger distance modulus) is necessary to match the cluster LF.  We
find that for a choice of [Fe/H]=-1.61, the Y$^2$ models must have age
of 15 (using the L98 color table) or 16 (using the G87 color table)
Gyr, respectively, to provide an adequate description of the M12 LF.
To retain an adequate fit with a younger age of 13 Gyr and metallicity
[Fe/H]=-1.61, the distance modulus would have to be larger that we
determined by more than $2\sigma$. This should once again emphasize the 
importance of renewed attention to $T_{\mbox{eff}}$-color transformations.

\section{Conclusions}

In this paper we have presented the $VI$ luminosity functions (LFs)
and $BVI$ color-magnitude diagrams (CMDs) of the Galactic globular cluster
(GGC) M12 from wide-field CCD photometry. Given constraints on the
cluster age, metallicity, distance modulus, and reddening we compare
our data to three sets of theoretical stellar evolution models for
metal-poor, $\alpha$-enhanced, low mass stars.  We find that neither
the \citet{bv01} nor the \citet{ki02} models are able to adequately
describe the SGB region of the M12 LF.  While we find no statistically
significant excesses of stars, the observed SGB LF has a noticeably
different slope than predicted.  We find the theoretical description
of the SGB region of the cluster LF to be sensitive to the selection
of color-$T_{\mbox{eff}}$ transformations, and to a lesser degree to age and
metallicity.  On the other hand, we find agreement between the
observed and predicted numbers of MS-to-RGB stars; M12 does not appear
to have an excess of RGB stars compared to theory.  In the context of the 
Langer et al. (2000) claim that extremely blue (``second parameter'')
clusters are explained by deep mixing (during the RGB phase) and
resulting envelope helium enrichment, the M12 LF should have shown this RGB
excess.  In contrast to this, we find the LF to be similar to that of 
M3 [$(B-R)/(B+V+R)=0.08$; Lee et al.(1994)], another cluster of nearly
identical metallicity that does not shown an excess of RGB stars. We find
in our analysis that regardless of the differences in input
physics in these two models, the adoption of the ZW metallicity scale is
incompatible with observations. Assuming a metallicity for M12 on the ZW
scale, adequate theoretical description of both the CMD and LF data would
require either (1) a model older than recent estimates of the age of
the universe (see below), or (2) a distance modulus that would be
$2-3\sigma$ larger than we determined from subdwarf fitting.

Analysis of the WMAP experimental data \citep{be03,sp03} has now
placed new restrictions to the age of the universe, and hence the
possible ages of GGCs.  Taking into account a GGC formation timescale
of $\sim1$ Gyr this implies that the possible ages of GGCs can be no
larger than $\sim13$ Gyr.  We find that the models of \citet{bv01}
require the use of a 16 Gyr model in order to account for the
observed properties of the M12 LF given the metallicity of M12 on the
ZW scale.  A 14 Gyr model would provide a similar fit but require the
use of a distance modulus that is $\sim0.15$ greater (just above our
$1\sigma$ error) than the value we derived from subdwarf fitting.
While comparisons between observations and the models of \citet{bv01}
have been shown to prefer the ZW metallicity scale \citep{bv01,va02},
use of the CG scale with these models still implies an age for M12 of
13-14 Gyr. In contrast, the theoretical models of \citet{ki02} and
\citet{va02} imply a cluster age of $12-13$ Gyr given the metallicity
of M12 on the CG scale, in better agreement with the WMAP upper
limit.  We attribute this to the use of He diffusion and
gravitational settling (only VandenBerg et al. 2002), although
uncertainties due to color-$T_{\mbox{eff}}$ transformations are still of 
comparable importance. Previous work (see $\S1$) has already shown
the input of diffusion tends to reduce the cluster ages by $0.5-2$
Gyr, and therefore provides the simplest explanation of our
observations.  Clearly for gravitational settling to become a
necessary part of stellar evolution models, confirmation of this age 
reduction and consistency with LF observations will be needed for
other clusters.  As a consequence, the reduction of systematic errors
in the distance modulus determination (and therefore cluster
metallicity and reddening) will be crucial to this analysis
\citep{gr03}. Further observations of stellar surface conditions,
such as the $^7$Li Spite plateau \citep{sp82,ry99} and metal abundance
variations in the GGC evolved populations, will also help to place
crucial constraints on gravitational settling as well \citep{va02,ch01}.

\acknowledgements

The authors wish to thank E. Brocato and K. von Braun for the generous use 
of their respective data sets.  We would also like to thank D. A. 
VandenBerg for the use of the Richard \& VandenBerg theoretical
isochrones and S. Yi for help on the use of the Y$^2$ luminosity
functions.  The authors wish to thank the referee R. Buonanno for his
insightful suggestions and criticisms.  This research has made use of the 
SIMBAD database, operated at CDS, Strasbourg, France.  This research was 
supported by National Science Foundation under grant AST 00-98696 to ELS 
and MB.

\clearpage

\clearpage


\begin{deluxetable}{ccccccc}
\tablewidth{0pt}
\tablecaption{Median residuals for comparison samples}
\tablehead{\colhead{Comparison} & \colhead{$B$}
& \colhead{$V$} & \colhead{$I$} & \colhead{($V-I$)} & \colhead{$(B-V)$} & \colhead{$N$}} 
\startdata
VB02 & \nodata & $-0.056\pm0.001$ & $-0.063\pm0.001$
& $0.006\pm0.002$ & \nodata & $8731$\\
R00 & \nodata & $-0.021\pm0.003$ & $-0.034\pm0.003$
& $0.015\pm0.002$ & \nodata & $2962$\\
B96 & $-0.048\pm0.011$ & $-0.021\pm0.008$ &
\nodata & \nodata & $-0.028\pm0.006$ & $1360$\\
S00 & $-0.010\pm0.009$ & $-0.022\pm0.005$ &
\nodata & \nodata & $0.009\pm0.010$ & $67$\\
\enddata
\label{comp}
\end{deluxetable}

\begin{deluxetable}{ccc}
\tablewidth{0pt}
\tablecaption{M12 [V,(B-V)] Fiducial Points}
\tablehead{\colhead{$V$} & \colhead{$(B-V)$} & \colhead{$N$}}
\startdata
19.8758 & 0.7487 & 572\\
19.7758 & 0.7470 & 557\\
19.6758 & 0.7334 & 609\\
19.5758 & 0.7285 & 548\\
19.4758 & 0.7095 & 555\\
19.3758 & 0.6999 & 602\\
19.2758 & 0.6989 & 574\\
19.1758 & 0.6892 & 536\\
19.0758 & 0.6832 & 567\\
18.9758 & 0.6735 & 522\\
\enddata
\label{bvfid}
\tablecomments{The complete version of this table 
is in the electronic edition of the Journal. The printed 
edition contains only a sample.}
\end{deluxetable}

\begin{deluxetable}{ccc}
\tablewidth{0pt}
\tablecaption{M12 [V,(V-I)] Fiducial Points}
\tablehead{\colhead{$V$} & \colhead{$(V-I)$} & \colhead{$N$}}
\startdata
20.2758 & 1.0705 & 371\\
20.1758 & 1.0502 & 419\\
20.0758 & 1.0322 & 529\\
19.9758 & 1.0180 & 534\\
19.8758 & 0.9978 & 572\\
19.7758 & 0.9800 & 557\\
19.6758 & 0.9700 & 609\\
19.5758 & 0.9478 & 548\\
19.4758 & 0.9352 & 555\\
19.3758 & 0.9227 & 602\\
\enddata
\label{vifid}
\tablecomments{The complete version of this table 
is in the electronic edition of the Journal. The printed 
edition contains only a sample.}
\end{deluxetable}

\begin{deluxetable}{ccc}
\tablewidth{0pt}
\tablecaption{M12 [V,(V-I)] Fiducial Points from \citet{vb02}}
\tablehead{\colhead{$V$} & \colhead{$(V-I)$} & \colhead{$N$}}
\startdata
22.1750 & 1.5157 & 254\\
22.0250 & 1.4734 & 379\\
21.8750 & 1.4278 & 437\\
21.7250 & 1.3946 & 421\\
21.5750 & 1.3405 & 397\\
21.4250 & 1.3099 & 405\\
21.2750 & 1.2608 & 364\\
21.1250 & 1.2353 & 382\\
20.9750 & 1.1951 & 371\\
20.8250 & 1.1625 & 362\\
\enddata
\label{vbfid}
\tablecomments{The complete version of this table 
is in the electronic edition of the Journal. The printed 
edition contains only a sample.}
\end{deluxetable}

\begin{deluxetable}{cccccccccccc}
\rotate
\tablewidth{0pt}
\tablecaption{Metal-Poor Subdwarfs with Well-Measured Parallaxes}
\tablehead{\colhead{HIC} & \colhead{HD/Gliese} & \colhead{$E(B-V)$} & \colhead{$V$} & \colhead{$\pi$ (mas)} & \colhead{$\sigma_{\pi}/\pi$} & \colhead{$M_V$} & \colhead{$\sigma_{M_V}$} & \colhead{\vi} & \colhead{[Fe/H]}  & \colhead{$\delta(V-I)$} & \colhead{$(V-I)_o$}}
\startdata
\cutinhead{Subdwarfs Used in MS Fit}
$38541$ & $ 64090$ & $ 0.000$ & $8.276$ & $35.29$ & $0.029$ & $6.01$ & $0.06$ & $0.771$ & $-1.48$ & $0.006$ & $0.801$\\
$57939$ & $ 103095$ & $0.000$ & $6.422$ & $109.21$ & $0.007$ & $6.61$ & $0.02$ & $0.891$ & $-1.24$ & $-0.005$ & $0.903$\\
$74234$ & $ 134440$ & $0.005$ & $9.418$ & $33.68$ & $0.050$ & $7.03$ & $0.11$ & $1.000$ & $-1.28$ & $0.014$ & $1.018$\\
$74235$ & $ 134439$ & $0.005$ & $9.052$ & $34.14$ & $0.040$ & $6.70$ & $0.08$ & $0.913$ & $-1.30$ & $0.004$ & $0.932$\\
$24316$ & $ 34328$ & $ 0.003$ & $9.436$ & $14.55$ & $0.069$ & $5.21$ & $0.15$ & $0.647$ & $-1.44$ & $-0.005$ & $0.673$\\
$98020$ & $ 188510$ & $0.001$ & $8.830$ & $25.32$ & $0.046$ & $5.83$ & $0.10$ & $0.753$ & $-1.37$ & $0.009$ & $0.774$\\
\cutinhead{Subdwarfs Eliminated from MS Fit}
$46120$ & Gl 345 & $0.012$ & $10.089$ & $16.46$ & $0.060$ & $6.14$ & $0.13$ & $0.728$ & $-1.75$ & $-0.042$ & $0.774$\\
$70681$ & $126681$ & $-0.001$ & $9.302$ & $19.16$ & $0.075$ & $5.66$ & $0.17$ & $0.727$ & $-0.90$ & $-0.037$ & $0.702$\\
$100568$ & $193901$ & $0.003$ & $8.644$ & $22.88$ & $0.054$ & $5.41$ & $0.11$ & $0.678$ & $-1.00$ & $-0.039$ & $0.664$\\
$67655$ & $120559$ & $0.020$ & $7.918$ & $40.02$ & $0.025$ & $5.92$ & $0.05$ & $0.755$ & $-0.95$ & $-0.044$ & $0.734$\\
$104659$ & $201891$ & $0.003$ & $7.367$ & $28.26$ & $0.036$ & $4.61$ & $0.08$ & $0.656$ & $-0.97$ & $0.016$ & $0.638$\\
$100792$ & $194598$ & $0.003$ & $8.335$ & $17.94$ & $0.069$ & $4.56$ & $0.15$ & $0.629$ & $-1.02$ & $-0.002$ & $0.616$\\
$18915$ & $ 25329$ & $ 0.000$ & $8.502$ & $54.14$ & $0.020$ & $7.17$ & $0.04$ & $1.007$ & $-1.69$ & $0.026$ & $1.06$\\
\enddata
\label{sdlistvi}
\end{deluxetable}


\begin{deluxetable}{cccc}
\tablewidth{0pt}
\tablecaption{M12 $V$ Band Luminosity Function}
\tablehead{\colhead{$V$} & \colhead{log $N$} &
\colhead{$\sigma_{high}$} & \colhead{$\sigma_{low}$}}
\startdata
12.065 & 0.1388 & 0.2323 & 0.5333\\
12.515 & 0.1388 & 0.2323 & 0.5333\\
12.665 & 0.1387 & 0.2323 & 0.5333\\
12.815 & 0.1390 & 0.2323 & 0.5333\\
12.965 & 0.1385 & 0.2323 & 0.5333\\
13.115 & 0.4402 & 0.1761 & 0.3010\\
13.415 & 0.3151 & 0.1979 & 0.3740\\
13.565 & 0.4409 & 0.1761 & 0.3010\\
13.715 & 0.6160 & 0.1487 & 0.2279\\
13.865 & 0.3151 & 0.1979 & 0.3740\\
\enddata
\label{volf}
\tablecomments{The complete version of this table
is in the electronic edition of the Journal. The printed
edition contains only a sample.}
\end{deluxetable}

\begin{deluxetable}{cccc}
\tablewidth{0pt}
\tablecaption{M12 $I$ Band Luminosity Function}
\tablehead{\colhead{$I$} & \colhead{log $N$} &
\colhead{$\sigma_{high}$} & \colhead{$\sigma_{low}$}}
\startdata
10.125 & -.1946 & 0.3011 & 1.0000\\
10.425 & 0.1065 & 0.2324 & 0.5338\\
10.725 & -.1946 & 0.3010 & 1.0000\\
10.875 & -.1946 & 0.3010 & 1.0000\\
11.025 & 0.5044 & 0.1606 & 0.2576\\
11.175 & -.1946 & 0.3010 & 1.0000\\
11.325 & 0.2826 & 0.1980 & 0.3742\\
11.625 & 0.2826 & 0.1980 & 0.3741\\
11.775 & 0.2826 & 0.1980 & 0.3742\\
11.925 & 0.1065 & 0.2323 & 0.5334\\
\enddata
\label{iolf}  
\tablecomments{The complete version of this table
is in the electronic edition of the Journal. The printed
edition contains only a sample.}
\end{deluxetable}

\begin{deluxetable}{lccc}
\tablecolumns{4}
\tablewidth{0pt}
\tablecaption{Input physics\,\tablenotemark{a}}
\tablehead{\colhead{} & \multicolumn{3}{c}{$\alpha$-enhanced Models} \\
\cline{2-4}\\
\colhead{Parameter} & \colhead{Y$^2$} & \colhead{BV} & \colhead{Richard \& VandenBerg}}
\startdata
Solar mixture & GN93 & AG89/G90,G91 & GN93\\
Initial He abundance & $Y_p=0.233$ & $Y_p=0.237$ & $Y_p=0.237$\\
Reaction Rates & BP92 & BP92 & BP92\\
Equation of State & OPAL96 (R96) & see Appendix of V00 & E73/CD92\\
Opacity & OPAL96 (RI95,IR96) & OPAL92 (RI92) & OPAL96\\
Low-Temperature Opacity & AF94 & AF94 & P94\\
$\alpha_{\textsc{\textrm{mlt}}}$ & $1.74$ & $1.89$ & $1.69$\\
Gravitational Settling & He diffusion (T94) & none & Yes (see T98)\\
Radiative Acceleration & none & none & R98\\
\enddata
\tablenotetext{a}{Abbreviations to references are as follows:
 BP92=\citet{bp92}; RI95=\citet{ro95}; IR96=\citet{ig96};
 AF94=\citet{al94}; T94=\citet{th94}; RI92=\citet{ro92}; R96=\citet{rog96};
 GN93=\citet{gr93}; AG89=\citet{an89}; G90=\citet{gr90};
 G91=\citet{gr91}; V00=\citet{va00}; R98=\citet{ri98}; E73=\citet{eg73};
 CD92=\citet{cd92}; P94=\citet{pr94}; T98=\citet{tu98}
}
\label{physics}
\end{deluxetable}
\end{document}